\documentclass[draftclsnofoot, 12pt, onecolumn]{IEEEtran}

\usepackage[english]{babel}
\usepackage{amsmath,amssymb}
\usepackage{times}
\usepackage{relsize}
\usepackage{graphicx}
\usepackage{url}
\usepackage{booktabs}
\usepackage{multirow}
\usepackage{mparhack}
\usepackage{subfigure}
\usepackage{authblk}
\usepackage{amsthm}
\usepackage{bm}
\usepackage{algorithm}
\usepackage[noend]{algorithmic}

\usepackage{array}
\usepackage{cite}
\usepackage{eqparbox}
\usepackage{mdwmath}

\usepackage{epsfig}
\usepackage{epstopdf}
\usepackage{xcolor}

\newtheorem{theorem}{Theorem}
\newtheorem{lemma}[theorem]{Lemma}

\newtheorem{corollary}[theorem]{Corollary}

\newcounter{exam}
\newenvironment{remark}[1][Remark.]{\begin{trivlist}
\item[\hskip \labelsep {\bfseries #1}]}{\end{trivlist}}

\DeclareFontFamily{OT1}{pzc}{}
\DeclareFontShape{OT1}{pzc}{m}{it}%
              {<-> s * [1.1] pzcmi7t}{}
\DeclareMathAlphabet{\mathpzc}{OT1}{pzc}%
                                 {m}{it}

\def\CN{{\mathcal N}}

\def\CJ{{\mathcal J}}

\def\CK{{\mathcal K}}

\def\CU{{\mathcal U}}

\newcommand{\Ct}{\mathpzc{t}}

\def\OP{{JPCAP}}
\def\WOP{{W-JPCAP}}
\def\SOP{{R-JPCAP}}

\def\ALG{{LDDP}}
\def\DP{{TSDP}}

\title{Power and Channel Allocation for Non-orthogonal 
Multiple Access in 5G Systems: Tractability and Computation}

\author[1]{Lei Lei}
\author[1,3]{Di Yuan}
\author[2]{Chin Keong Ho}
\author[2]{Sumei Sun}
\affil[1]{{\small Department of Science and Technology, Link{\"o}ping University, Sweden}}
\affil[2]{{\small Institute for Infocomm Research (I$^2$R), A$^*$STAR, Singapore}}
\affil[3]{{\small Institute for Systems Research, University of Maryland, College Park, USA}}
\affil[ ]{\em{{\small Emails: \{lei.lei; di.yuan@liu.se\},  \{hock; sunsm\}@i2r.a-star.edu.sg}}}

\begin{document}

\date{}

\maketitle

\begin{abstract}

Network capacity calls for significant increase for 5G cellular
systems.  A promising multi-user access scheme, non-orthogonal
multiple access (NOMA) with successive interference cancellation
(SIC), is currently under consideration.  In NOMA, spectrum efficiency
is improved by allowing more than one user to simultaneously access
the same frequency-time resource and separating multi-user signals by
SIC at the receiver.  These render resource allocation and
optimization in NOMA different from orthogonal multiple access in 4G.
In this paper, we provide theoretical insights and algorithmic
solutions to jointly optimize power and channel allocation in NOMA.
For utility maximization, we mathematically formulate NOMA resource
allocation problems.  We characterize and analyze the problems'
tractability under a range of constraints and utility functions.  For
tractable cases, we provide polynomial-time solutions for global
optimality.  For intractable cases, we prove the NP-hardness and
propose an algorithmic framework combining Lagrangian duality and
dynamic programming (LDDP) to deliver near-optimal solutions.  To
gauge the performance of the obtained solutions, we also provide
optimality bounds on the global optimum.  Numerical results
demonstrate that the proposed algorithmic solution can significantly
improve the system performance in both throughput and fairness over
orthogonal multiple access as well as over a previous NOMA resource
allocation scheme.
\end{abstract}


\section{Introduction}

Orthogonal multi-user access (OMA) techniques are used in 4G long term
evolution (LTE) and LTE-Advanced (LTE-A) networks, e.g., orthogonal
frequency division multiple access (OFDMA) for downlink and
single-carrier frequency division multiple access (SC-FDMA) for uplink
\cite{ofdma,scfdma}.  In OMA, within a cell, each user has exclusive
access to the allocated resource blocks.  Thus, each subchannel or
subcarrier can only be utilized by at most one user in every time
slot.  OMA avoids intra-cell interference, and enables single-user
detection/decoding and simple receiver design. However, by its nature,
orthogonal channel access is becoming a limiting factor of spectrum
efficiency.

In the coming decade, the mobile data traffic is expected to grow
thousand-fold \cite{5gbe, vtm}.  Accordingly, the network capacity
must dramatically increase for 5G systems.  Capacity scaling for 5G is
enabled by a range of techniques and schemes, e.g., cell
densification, utilization of unlicensed spectrum, and advanced radio
access schemes.  New multi-user access schemes have been investigated
as potential alternatives to OFDMA and SC-FDMA
\cite{5gbe,vtm,aim5g}. A promising scheme is the so-called non-orthogonal
multiple access (NOMA) with successive interference cancellation (SIC)
\cite{higuchi}. Unlike interference-avoidance multiple access schemes, e.g., OFDMA,
multiple users in NOMA can be assigned to the same frequency-time
resource so as to improve spectrum efficiency \cite{higuchi}.  On one
hand, this results in intra-cell interference among the multiplexed
users.  On the other hand, some of the interfering signals in NOMA can
be eliminated by multi-user detection (MUD) with SIC at the receiver
side.  To enable this process, more advanced receiver design and
interference management techniques are considered to be the key
aspects in 5G networks
\cite{vtm}.

\subsection{Related Works}

From an information theory perspective, under the assumption of
simultaneous multi-user transmission via superposition coding with
SIC, capacity region and duality analysis have been studied in
\cite{duality}.  The authors of \cite{overview} have provided an
analysis of implementing interference cancellation in cellular
systems.  Towards future 5G communication systems, some candidate
access schemes are under investigation in recent research activities,
e.g., sparse code multiple access (SCMA) \cite{5gbe}, and NOMA
\cite{higuchi}.

In \cite{nomainf}, the authors studied the capacity region for NOMA.
In \cite{system}, by assuming predefined user groups for each
subchannel, a heuristic algorithm for NOMA power allocation in
downlink has been proposed, and system-level simulations have been
conducted.  The authors in \cite{globecom} considered a sum-rate
utility maximization problem for dynamic NOMA resource allocation.  In
\cite{random}, outage performance of NOMA has been evaluated.  The
authors derived the ergodic sum rate and outage probability to
demonstrate the superior performance of NOMA with fixed power
allocation.  In \cite{fairness}, fairness considerations and a max-min
fairness problem for NOMA have been addressed.  The fairness in NOMA
can be improved via using and adapting the so-called power allocation
coefficients.  For uplink NOMA, the authors in \cite{uplinksic}
provided a suboptimal algorithm to solve an uplink scheduling problem
with fixed transmission power.  In \cite{uplink12}, a weighted
multi-user scheduling scheme is proposed to balance the total
throughput and the cell-edge user throughput.  In \cite{uplink14}, the
authors proposed a greedy-based algorithm to improve the throughput in
uplink NOMA.  In \cite{DiFaPo15}, the authors studied and evaluated
user grouping/pairing strategies in NOMA. It has been shown that, from
the outage probability perspective, it is preferable to multiplex
users of large gain difference on the same subcarrier.  We also remark that there are other setups of SIC than that considered
in NOMA.  An example is the interference channel in which common
information is transmitted for partial interference cancellation, for which
Etkin~et~al. \cite{EtTsWa08} provided an analysis of the resulting
capacity region and trade-off from an information theory
perspective.

Apart from investigation of NOMA performance in cellular networks,
from a general optimization perspective, the complexity and
tractability analysis of NOMA resource allocation is of significance.
Here, tractability for an optimization problem refers to whether or
not any polynomial-time algorithm can be expected to find the global
optimum \cite{ofdma}.  Tractability results for resource allocation in
OMA and interference channels have been investigated in a few existing
works, e.g., \cite{ofdma,
yafeng,limit} for OFDMA, \cite{scfdma} for SC-FDMA, and \cite{luo, limit}  for
interference channel.  
For NOMA, to the best of our
knowledge, no such study is available in the existing literature.

\subsection{Contributions}

In spite of the existing literature of performance evaluation for
NOMA, there is lack of a systematic approach for NOMA resource
allocation from a mathematical optimization point of view.  The
existing resource allocation approaches for NOMA are typically carried
out with fixed power allocation \cite{random, uplinksic, uplink14},
predefined user set for subchannels \cite{system}, or parameter tuning
to improve performance, e.g., updating power allocation coefficients
\cite{fairness}.  Moreover, compared with OMA, NOMA allows multi-user
sharing on the same subchannel, thus provides an extra dimension to
influence the performance in throughput and fairness.  However, how to
balance these two key performance aspects in power and channel
allocation is largely not yet studied in the literature.  In addition,
little is known on the computational complexity and tractability of
NOMA resource allocation.

In this paper, the solutions of joint channel and power allocation for
NOMA are subject to systematic optimization, rather than using
heuristics or ad-hoc methods.  To this end, we formulate, analyze, and
solve the power and channel optimization problem for downlink NOMA
systems, taking into account practical considerations of fairness and
SIC.  We present the following contributions.  First, for maximum
weighted-sum-rate (WSR) and sum-rate (SR) utilities, we formulate the
joint power and channel allocation problems (\OP{}) mathematically.
Second, we prove the NP-hardness of \OP{} with WSR and SR utilities.
Third, we identify tractable cases for \OP{} and provide the
tractability analysis.  Fourth, considering the intractability of \OP{}
in general, we propose an algorithmic framework based on Lagrangian
duality and dynamic programming to facilitate problem solving.  Unlike
previous works, our approach contributes to delivering near-optimal
solutions, as well as performance bounds on global optimum to
demonstrate the quality of our near-optimal solutions. We
use numerical results to illustrate the significant performance
improvement of the proposed algorithm over existing NOMA and OFDMA
schemes. 

Our work extends previous study of user grouping in NOMA.  In
\cite{DiFaPo15}, the number of users to be multiplexed on a subcarrier
is fixed, and performance evaluation consists of rule-based 
multiplexing policies. In our case, for each subcarrier, the number of
users and their composition are both output from solving an
optimization problem. Later in Section~VII, results of optimized
subcarrier assignment and user grouping will be presented for
analysis.
The current paper extends our previous study
\cite{globecom} in several dimensions. The extensions consist of the
consideration of the WSR utility metric, a significant amount of
additional theoretical analysis of problem tractability, the
development of the performance bound on global optimality, as well as
the consideration of user fairness in performance evaluation.

The rest of the paper is organized as follows. Section II gives the
system models for single-carrier and multi-carrier NOMA cellular
systems. Section III formulates \OP{} for WSR utility and provides
complexity analysis.  Section IV analyzes the tractability for special
cases of \OP{}.  In Section V, we provide the tractability analysis
for relaxations of \OP{}.  Section VI proposes an algorithmic
framework for \OP{}.  Numerical results are given in Section
VII. Conclusions are given in Section VIII.

\section{System Model}

\subsection{Basic Notation}

We consider a downlink cellular system with a base station (BS)
serving $K$ users.  The overall bandwidth $B$ is divided into $N$
subchannels, each with bandwidth $B/N$.  Throughout the paper, we
refer to subchannel interchangeably with subcarrier.  We use $\CK$ and
$\CN$ to denote the sets of users and subchannels, respectively, and
$g_{kn}$ to denote the channel gain between the BS and user $k$ on
subcarrier $n$.  Let $p_{kn}$ be the power allocated to user $k$ on
subcarrier $n$.  A user $k$ is said to be multiplexed on a subchannel
$n$, if and only if $p_{kn}>0$.  The power values are subject to
optimization.  At the receiver, each user equipment has MUD
capabilities to perform multi-user signal decoding \cite{iccs}.  With
SIC, some of the co-channel interference will be treated as decodable
signals instead of as additive noise.

\subsection{NOMA Systems}
\label{sec:noma}

To ease the presentation of the system model, for the moment let us
consider the case that all the $K$ users can multiplex on each
subcarrier $n$ in a multi-carrier NOMA system (MC-NOMA) at downlink.
For each subcarrier $n$, we sort the users in set $\CK$ in the
descending order of channel gains, and use bijection $b_n(k)$: $\CK
\mapsto \{1, 2, \dots, K\}$ to represent this order, where $b_n(k)$ is
the position of user $k \in \CK$ in the sorted sequence.  For our
downlink system scenario, in \cite{tse} (Chapter~6.2.2) it is
shown that, with superposition coding, user $k$ with better channel
gain $g_{kn}$ can decode the signal of user $h$ with worse channel
gain $g_{hn}$, and this is not constrained by the specific power
allocation.  Thus receiver $k$ is able to perform SIC, by subtracting
the re-encoded signal intended for receiver $h$ from the composite
signal, and thereby obtaining an increase in the
signal-to-interference-plus-noise ratio (SINR).  
Thus, user $k$ on
subcarrier $n$, before decoding its signal of interest, first decodes
the received interfering signals intended for the users $h \in \CK
\backslash \{k\}$ that appear later in the sequence than $k$, i.e.,
$b_n(h)>b_n(k)$.  The interfering signals with order $b_n(h)<b_n(k)$
will not be decoded and thus treated as noise.  Hence, the
interference after SIC for user $k$ on subcarrier $n$ is
$\sum_{\substack{h \in \CK \backslash \{k\}: b_n(h)<b_n(k)}} \limits
\hspace{-5mm} p_{hn} g_{kn}$, $\forall k \in \CK, \forall n \in
\CN$. If there are users having the same channel gain, then SIC
applies following the principle in \cite{tse} (Chapter~6.2.2),
provided that an ordering of the users is given.  From the discussion,
the SINR of user $k$ on subcarrier $n$ is given below.
\begin{equation}
\label{eq:sinr}
{\text{SINR}}_{kn}= \frac{p_{kn}g_{kn}} {\sum_{\substack{h \in \CK \backslash \{k\}:  b_n(h)<b_n(k)}} \limits p_{hn} g_{kn} +\eta}
\end{equation}
The noise power is denoted by $\eta$, which equals the product of the
power spectral density of white Gaussian noise and the subcarrier
bandwidth.  The rate of each user in NOMA is determined by the user's
SINR after SIC.  Thus, the achievable rate of user $k$ on subcarrier
$n$ is $R_{kn} =\log(1+{\text{SINR}_{kn}})$ nat/s with normalized
bandwidth $\frac{B}{N}=1$.

For single-carrier NOMA systems (SC-NOMA), we omit the subcarrier
index.  For convenience, the users $k \in \{1, \dots, K\}$ in SC-NOMA
are defined in the descending order of channel gains, where $g_1 \geq
g_2 \geq ,\dots, \geq g_K$.  Thus the user index also represents its
position in the sequence, and user $k$ is able to decode the signal of
user $h$ if $k < h$.  We define ${\text{SINR}}_{k}=
\frac{p_{k}g_{k}}{\sum_{\substack{h \in \CK \backslash \{k\}:
h<k}}p_{h} g_{k} +\eta}$, $\forall k \in \CK$.  The achievable rate of
user $k$ is $R_{k}$ = $\log(1+{\text{SINR}_{k}})$ nat/s with
normalized bandwidth $B=1$.  For illustration, consider a two-user
case with $g_1 >g_{2}$ in SC-NOMA.  Following the system model, user
$1$ is capable of decoding the interfering signal from user $2$.  For
user $1$, before decoding its own signal, the interference from user
2's signal is first decoded and removed, resulting in user 1's
achievable rate $\log(1+\frac{p_1g_1}{\eta})$.  For user $2$, no SIC
takes place, hence the achievable rate is
$\log(1+\frac{p_2g_2}{p_1g_2+\eta})$.

We use $\CU_n$ as a generic notation for the set of users multiplexed
on subchannel $n$ for MC-NOMA.  For SC-NOMA, the corresponding entity
is denoted by $\CU$. We use $M$, $1 \leq M \leq K$, to denote the
maximum number of multiplexed users on a subcarrier.  The reason of
having this parameter is to address complexity considerations of
implementing MUD and SIC.  In NOMA, the system complexity increases by $M$, because a user device needs
to decode up to $M$ signals.  The setting of $M$ depends on receiver's
design complexity and signal processing delay for SIC \cite{vtm,
overview}. For practical implementation, $M$ is typically smaller than
$K$. However, our optimization formulations and the solution
algorithm are applicable to any value of $M$ between one and $K$.

Two utility functions, WSR and SR, are considered in this paper.  The
WSR utility is denoted by ${f}_{\text W}= \sum_{k \in \CK}w_{k}\sum_{n
\in \CN}R_{kn}$, where $w_k$ is the weight coefficient of user $k \in \CK$.
Clearly, the selection of the weights has strong influence on the
resource allocation among the users. In general, the weights can be
used to steer the resource allocation towards various goals, such as
to implement service class priority of users, and fairness (e.g., a
user with averagely poor channel receives higher weight).  In our
work, the algorithmic approach is applicable without any assumption of
the specific weight setting. For performance evaluation, we set the
weights following proportional fairness.  That is, for one time slot,
a user's weight is set to be the reciprocal of the average user rate
prior to the current time slot \cite{uplinksic}.  
As a result, the
resource allocation will approach proportional fairness over time.
The SR utility, a special case of WSR, is defined
as ${f}_{\text R}= \sum_{k \in \CK}\sum_{n \in \CN}R_{kn}$.  The term
of SR utility is used interchangeably throughput in this paper.  For
both SR and WSR, SIC with superposition coding \cite{tse} applies to
the users multiplexed on the same subcarrier. As was discussed
earlier, the decoding does not rely on assuming specific, a priori
constraint on the power allocation among the users.

\section{Joint Power and Channel Allocation}

In this section, we formulate \OP{} using WSR utility for MC-NOMA.  We
use \WOP{} to denote the optimization problem.  In general, \OP{}
amounts to determining which users should be allocated to which
subcarriers, as well as the optimal power allocation such that the
total utility is maximized.  In the following we define the variables
and formulate \WOP{} as $P1_{\text WSR}$ below, where all
$p$-variables and $x$-variables are collected in vectors $\bm{p}$ and
$\bm{x}$, respectively.

\vspace{-8mm}
$$
\begin{array}{ll}
\hspace{-6cm}p_{kn}& = \ \textrm{allocated power to user $k$ on subcarrier $n$.}
\end{array}
$$
$x_{kn}$ = 
    $\left\{
      \begin{array}{l l}      
        1 & $if user $k$ is multiplexed on subcarrier $n$$,\ $i.e., $p_{kn}>0$$, \\
        0 & $otherwise$. \\
      \end{array}
    \right.$

\begin{subequations}
\label{eq:p1}
\begin{align}
\text{$P1_{\text WSR}$:}  
\label{eq:p1obj}
  \qquad & \hspace{-4mm} \max_{\bm{x}, \bm{p}} \limits\ 
  \sum_{k \in \CK}w_k\sum_{n \in \CN} R_{kn}x_{kn} \qquad &  \\
\label{eq:p11}
\qquad &   \text{s.t.}  \ \ 
\sum_{k \in \CK}\sum_{n \in \CN} p_{kn} \leq P_{tot}  &\\
\label{eq:p12}
\qquad &  
\hspace{7mm} \sum_{n \in \CN} p_{kn} \leq P_{k},  \ \forall k \in \CK &\\
\label{eq:p13}
\qquad &  
\hspace{7mm} \sum_{k \in \CK} x_{kn} \leq M,  \ \forall n \in \CN &
\end{align}
\end{subequations}

In $P1$, the objective (\ref{eq:p1obj}) is to maximize the WSR
utility, where $R_{kn}$ contains the $p$-variables, see
(\ref{eq:rkn}) below.
\begin{equation}
\label{eq:rkn}
R_{kn}=\log(1+\frac{p_{kn}g_{kn}} {\sum_{\substack{h \in \CK \backslash \{k\}: \\ b_n(h)<b_n(k)}} \limits p_{hn} g_{kn} +\eta}), \forall k \in \CK, \ \forall n \in \CN
\end{equation}
Constraints (\ref{eq:p11}) and (\ref{eq:p12}) are respectively imposed
to ensure that the total power budget and the individual power limit
for each user are not exceeded.  The per-user power limit $P_k$ is
introduced for practical considerations, such as regulatory
requirement on power towards a user device. Such a limit is also very
common in OMA (e.g., \cite{luo, limit}).  Constraints (\ref{eq:p13})
restrict the maximum number of multiplexed users on each subcarrier
to $M$.  We remark that the power allocation is represented by the
$p$-variables of which the values are subject to optimization, whereas
the power limits $P_{tot}$ and $P_k, k \in \CK$ are given entities.
Suppose $M$ users, say users $1, \dots M$, are allocated with positive
power on channel $n$. If $P_{tot} \geq \sum\limits_{k=1}^M P_k$
happens to hold, then all the $M$ users may be at their respective
power limits, i.e., setting $p_{kn}=P_k, k=1,\dots, M$, is feasible.
Otherwise, the $M$ users can still be allocated with positive power,
though not all of them can be at the power limits. Indeed, if user $k$
is allocated with power $p_{kn} > 0$ on channel $n$, then typically
$p_{kn} < P_k$ unless user $k$ is not allocated power on any other
channel than $n$. For the total power limit $P_{tot}$ to be
meaningful, one can assume $\sum_{k \in \CK} P_k > P_{tot}$ without
loss of generality, because otherwise the total power limit $P_{tot}$ is
not violated even if all users are allocated with their respective
maximum power, that is, (\ref{eq:p11}) becomes void and should be dropped. We do
not consider any further specific assumptions on the relation between
$P_{tot}$ and $P_k, k \in \CK$, to keep the generality of the system
model.

\begin{remark}
We do not explicitly impose the constraint that $p_{kn}>0$ if and only
if $x_{kn}=1$. This is because setting $p_{kn}>0$ and $x_{kn}=0$ is
clearly not optimal, by the facts that $p_{kn}>0$ will lead to 
rate degradation of other users due to the co-channel interference
(if there are other users on channel $n$), and that for user $k$,
$p_{kn}>0$ means power is consumed, but $x_{kn}=0$ means no benefit
as the rate in \eqref{eq:p1obj} becomes zero. $\Box$
\end{remark}

It can be observed that formulation $P1_{\text WSR}$ is non-linear and
non-convex.  The concavity of the objective function (\ref{eq:p1obj})
can not be established in general, because of the presence of the
binary $x$-variables and the product of $x$ and $p$.  Note that in
complexity theory, neither non-convexity nor non-linearity of a
formulation proves the problem's hardness, as a problem could be
inappropriately formulated.  Therefore, we provide formal hardness
analysis for \WOP{} below.

\begin{theorem}
\WOP{} is NP-hard.
\label{th:wnphard}
\end{theorem}  
\begin{IEEEproof}
We establish the result in two steps. First, we conclude that if $M=1$ in (\ref{eq:p13}),
\WOP{} is NP-hard, as it reduces to OFDMA subcarrier and power allocation,
for which NP-hardness  is provided in \cite{yafeng}
and \cite{limit}. For general MC-NOMA with $M >1$, we construct an instance of
\WOP{} and establish the equivalence between the instance and the OFDMA
problem considered in \cite{limit}.  We consider an instance of \WOP{}
with $K$ users, $N$ subcarriers, and $M=2$.  Let $\epsilon$ denote a
small value with $0<\epsilon<\frac{1}{e}^{KN}$.  The total power
$P_{tot}$ is set to $NKP_k$.  The power limit $P_k=1$ is uniform for
$\forall k \in \CK$, and the noise parameter $\eta=\epsilon$.  Among
the $K$ users, we select an arbitrary one, denoted by $\bar k \in
\CK$, and assign a dominating weight $w_{\bar k}=e^{KN}$ and channel
gain $g_{\bar kn}=1$ on all the subcarriers, whereas the other users'
weights and channel gains are $w_{k} = \epsilon$ and $g_{kn} \leq
\frac{1}{e}^{KN}$, $\forall k \in \CK \backslash \{\bar k\}$ and
$\forall n \in \CN$.  From above, the ratios $\frac{w_{\bar
k}}{w_{k}}$ and $\frac{g_{\bar kn}}{g_{kn}}$ are sufficiently large
such that allocating any power $p \leq P_{\bar k}$ to user $\bar k$ on
any subcarrier $n$, the utility $w_{\bar k}R_{\bar kn} > \max(\sum_{k
\in \CK \backslash \{\bar k\}}\sum_{n \in \CN}w_{kn}R_{kn})$ for using
the same power budget $p$, since $\sum_{k \in \CK \backslash \{\bar
k\}}\sum_{n \in \CN}w_{kn}R_{kn}$ is bounded by
$KNe^{-KN}\log(1+\frac{e^{-KN}p}{\epsilon})$, and $w_{\bar k}R_{\bar
kn}=e^{KN}\log(1+\frac{p}{\epsilon})$ is clearly greater than
$KNe^{-KN}\log(1+\frac{e^{-KN}p}{\epsilon})$.  Thus, allocating power
to user $\bar k$ rather than other users is preferable for maximizing
utility.

Due to the uniform gain $g_{\bar kn}$ and the dominating weight
$w_{\bar k}$ for user $\bar k$ on all channels, the optimal power
allocation for user $\bar k$ is to uniformly allocate an amount of
$\frac{P_{\bar k}}{N}$ to each subcarrier.  Then the remaining problem
is to allocate power $P_{tot}-P_{\bar k}=(NK-1)P_k$ to the remaining
$K-1$ users.  Every user $k \in \CK \backslash \{\bar k\}$ is still
subject to the user power constraint (\ref{eq:p12}).  Note that for
$M=2$, each subcarrier now can accommodate one extra user at most.
Compared to the OFDMA problem in \cite{limit}, \WOP{} has one extra
total power constraint, i.e., (\ref{eq:p11}), however, recall that
$P_{tot}$ is set to $NKP_k$, and for this value (\ref{eq:p11}) is in
fact redundant.  Therefore, a special case of \WOP{} with $M>1$ is
equivalent to the OFDMA problem in \cite{limit}, and the result
follows.
\end{IEEEproof}

\section{Tractability Analysis for Uniform Weights}
\label{sec:opf}

The hardness of \WOP{} could have stemmed from several sources, e.g.,
the structure of the utility function, discrete variables, non-concave
objective, and the constraints.  We start from investigating how the
weight in the utility function influences the problem's tractability.
The utility function can affect the computational complexity in
problem solving \cite{luo, yafeng}.  One example is that the SR
maximization problem with total power constraint in OFDMA is
polynomial-time solvable \cite{yafeng}.  With WSR utility, solving the
same problem is challenging \cite{weight}.  In this section, we
consider a special problem of \WOP{}, i.e., SR utility with uniform
weights for users.  We use \SOP{} to denote the optimization problem.
Intuitively, \SOP{} appears somewhat easier than \WOP{}, however, the
tractability of \SOP{}, for both SC-NOMA and MC-NOMA, is not known in
the literature.  In the following, analogously to \WOP{}, we formulate
\SOP{} in $P1_{SR}$.

\vspace{-8mm}
\begin{subequations}
\label{eq:p1s}
\begin{align}
\text{$P1_{\text SR}$:}  
\label{eq:p1sobj}
  \qquad & \max_{\bm{x}, \bm{p}} \limits \ 
  \sum_{k \in \CK}\sum_{n \in \CN} R_{kn}x_{kn} \qquad &  \\
\qquad &   \text{s.t.}  \ \ 
\text{(\ref{eq:p11}), (\ref{eq:p12}), (\ref{eq:p13})}
\end{align}
\end{subequations}

First, we provide structural insights for the optimal power
allocation for \SOP{} in SC-NOMA.

\begin{theorem}
\label{th:noskip}
For \SOP{} in SC-NOMA, 
the following hold at the optimum: \\
(a) Suppose $g_{k} \geq g_{h}$ for two users $k$ and $h$ with $k \neq h$, then  the optimal power allocation satisfies $p_{h} > 0$ only if $p_{k} > 0$. \\ 
(b) Up to $M$ consecutive users in descending order of channel gain are allocated with positive power.
\end{theorem}

\begin{IEEEproof}
As defined in the SC-NOMA system model, the SR utility function ${f}={f}(p_{1}, \dots, p_{K})_{\text R}$ reads:
\begin{equation}
\begin{aligned}
\label{logsum}
&\hspace{-1mm}
\log\hspace{-1mm}\left(\hspace{-1mm}1\hspace{-1mm}+\hspace{-1mm}\frac{p_1g_1}{\eta}\hspace{-1mm}\right)
\hspace{-1mm} +
\hspace{-1mm}\log\hspace{-1mm}\left(\hspace{-1mm}1 \hspace{-1mm} + \hspace{-1mm} \frac{p_2g_2}{p_1g_2\hspace{-1mm}+\hspace{-1mm}\eta}\hspace{-1mm}\right)\hspace{-1mm}+ ,\dots,  +
\log\hspace{-1mm}\left(\hspace{-1mm}1\hspace{-1mm}+\hspace{-1mm}\frac{p_Kg_K}{\sum_{\substack{ \\ h =1}}^{K-1} \limits \hspace{-1mm} p_{h} g_{K}\hspace{-1mm}+\hspace{-1mm}\eta}\hspace{-1mm}\right)    \notag \\ 
\end{aligned}
\end{equation}
\vspace{-0.2cm}
\begin{equation}
\begin{aligned}
= &\log{(p_1g_1+\eta)} - \log{\eta}+\log{((p_1+p_2)g_2+\eta)} -  \\
& \hspace{-3mm} \log{(p_1g_2+\eta)}+, \dots,+ \hspace{-1mm} \log{(\sum_{\substack{ \\ h =1}}^{K} \limits  p_{h} g_{K} \hspace{-1mm} +\hspace{-1mm} \eta)} \hspace{-1mm} - \hspace{-1mm} \log{(\sum_{\substack{ \\ h =1}}^{K-1} \limits p_{h} g_{K} \hspace{-1mm} + \hspace{-1mm} \eta)} \\
\notag
\end{aligned}
\vspace{-0.6cm}
\end{equation}  
Next, we consider the partial derivatives $\frac{\partial {
f}}{\partial p_1},\dots$, $\frac{\partial {f}}{\partial p_k},\dots,
\frac{\partial {f}}{\partial p_K}$ for each $p$-variable as shown in (\ref{pd}).

\vspace{-0.4cm}
\begin{equation}
\label{pd}
\left\{
\begin{aligned}
&\hspace{-1mm} \frac{\partial {f}}{\partial p_1} = \underbrace{ \frac{g_1}{p_1g_1\hspace{-1mm}+\hspace{-1mm}\eta}\hspace{-1mm}-\hspace{-1mm}\frac{g_2}{p_1g_2\hspace{-1mm}+\hspace{-1mm}\eta}}_{\geq 0} \hspace{-1mm}+ \hspace{-1mm}  \underbrace{\frac{g_2}{(p_1\hspace{-1mm}+\hspace{-1mm}p_2)g_2\hspace{-1mm}+\hspace{-1mm}\eta} -\frac{g_3}{(p_1\hspace{-1mm}+\hspace{-1mm}p_2)g_3\hspace{-1mm}+\hspace{-1mm}\eta} }_{\geq 0} \hspace{-1mm}+\\
& ,\dots,+ \underbrace{ \frac{g_{K-1}}{\sum_{\substack{ \\ h =1}}^{K-1} \limits p_{h} g_{K-1}\hspace{-1mm}+\hspace{-1mm}\eta} 
-\frac{g_{K}}{\sum_{\substack{ \\ h =1}}^{K-1} \limits  p_{h} g_{K}\hspace{-1mm}+\hspace{-1mm}\eta}}_{\geq 0}
+ \frac{g_K}{\sum_{\substack{ \\ h =1}}^{K} \limits p_{h} g_{K}\hspace{-1mm}+\hspace{-1mm}\eta}, \\ 
& \frac{\partial{f}}{\partial p_2},  \\
& \vdots \\
& \frac{\partial{f}}{\partial p_{K-1}}\hspace{-1mm}=\hspace{-1mm} \underbrace{ \frac{g_{K-1}}{\sum_{\substack{ \\ h =1}}^{K-1} \limits \hspace{-1mm}p_{h} g_{K-1}\hspace{-1mm}+\hspace{-1mm}\eta} 
-\frac{g_{K}}{\sum_{\substack{ \\ h =1}}^{K-1} \limits  \hspace{-1mm}p_{h} g_{K}\hspace{-1mm}+\hspace{-1mm}\eta}}_{\geq 0}
+ \frac{g_K}{\sum_{\substack{ \\ h =1}}^{K} \limits p_{h} g_{K}\hspace{-1mm}+\hspace{-1mm}\eta},  \\
& \frac{\partial {f}}{\partial p_K}=\frac{g_K}{\sum_{\substack{ \\ h =1}}^{K}  p_{h} g_{K}+\eta}  
\end{aligned} 
\right. 
\end{equation} 

For conclusion (a), since $g_{1} \geq g_{2} \geq, \dots, \geq g_{K}$
for the single subcarrier in SC-NOMA, it can be easily checked from
(\ref{pd}) that the partial derivatives $\frac{\partial {f}}{\partial
p_1} \geq \frac{\partial {f}}{\partial p_2} \geq \frac{\partial
{f}}{\partial p_3} \geq, \dots, \geq\frac{\partial {f}}{\partial
p_{K}} > 0$, irrespective of the power values.  In general, from the
partial derivatives, more utility will be obtained by increasing power
$p_k$ instead of $p_h$ if $k<h, \ \forall k , h \in \CK$.  If the user
with the best channel condition, i.e. user 1, has $p_{1} < P_1$, the
objective value ${f}(p_{1}, \dots, p_{K})_R$ can be improved by
shifting power from other users to user 1 until $p_{1}$ equals the
power limit $P_{1}$.  Statement (a) also implies that the optimal
power allocation will be in a consecutive manner, i.e., from user 1 to
user $M$, one by one, and the result of (b) follows.
\end{IEEEproof}

By Theorem~\ref{th:noskip}, for the SR utility, the users being
allocated positive power at optimum are consecutive in their gain
values, starting from the user with the best channel gain.  Note that
the SR utility maximizes the throughput, but has the issue of
fairness, which is addressed by the more general metric of weighted SR
(WSR) utility, where the weights are set according to the proportional
fairness policy. By Theorem \ref{th:noskip} and its proof, \SOP{} for
SC-NOMA can be optimally solved by the procedure given in
Algorithm~\ref{alg:poly}.  The users' power allocation is performed in
a consecutive manner, starting from user $k=1$, and assigning power
$\min(P_k, P_{tot})$ to user $k$ and updating $P_{tot}$ accordingly.
Algorithm \ref{alg:poly} is clearly of polynomial-time complexity,
resolving the tractability of \SOP{} in SC-NOMA, giving Corollary
\ref{th:sop} below.

\begin{corollary}
\label{th:sop}
\SOP{} for SC-NOMA is tractable, i.e., polynomial-time solvable.
\end{corollary}  

\begin{algorithm}[ht!]
\caption{Polynomial-Time Algorithm for \SOP{} in SC-NOMA }
\label{alg:poly}
\begin{algorithmic}[1]
\STATE Initialize $p_{k}^*=0, \ \forall k \in \CK$
\FOR{$k=1:M$}   
\STATE {$p_k^* \gets \min(P_k, P_{tot})$, $P_{tot}=P_{tot} - p_{k}^*$} \label{min}
\IF{$P_{tot}=0$}
\STATE {Break}
\ENDIF
\ENDFOR 
\STATE {{\bf Return}: Optimal power allocation $p_{1}^*, \dots, p_{M}^*$}
\end{algorithmic}
\end{algorithm}

Next, we analyze the computational complexity of \SOP{} in MC-NOMA.

\begin{theorem}
\label{the:sop}
\SOP{} for MC-NOMA is NP-hard.
\end{theorem}  
\begin{IEEEproof}
The proof is analogous to that of Theorem \ref{th:wnphard}.  For
general \SOP{} with $M >1$, we construct a special instance with $K$
users, $N$ subcarriers, $M=2$, $P_{tot}=NKP_k$, and uniform $P_k=1$.
We deploy a ``dominant user" $\bar k \in \CK$ in the instance, that
is, user $\bar k$ has the highest and uniform channel gain of $g_{\bar
kn}=1$ for all the subcarriers, whereas the channel gain of all the
other users and subcarriers is $g_{kn} \leq \frac{1}{e}^{KN}$,
$\forall k \in \CK \backslash \{\bar k\}, \ \forall n \in \CN$.  One
can observe that the ratio $\frac{g_{{\bar k}n}}{g_{kn}}$ for $\forall
n \in \CN$, $\forall k \in \CK \backslash \{\bar k\}$ has been set
sufficiently large, and from (\ref{pd}), the partial derivative of
user $\bar k$ satisfies $\frac{\partial {f}}{\partial p_{\bar kn}}>
\frac{\partial {f}}{\partial p_{kn}}$ for $\forall k \in \CK
\backslash \{\bar k\}$ on each $n$, irrespective of the power values.
Then the statement (a) in Theorem~\ref{th:noskip} is valid for any
subcarrier $n$ in this instance, that is, if $\bar k$ is multiplexed
on subcarrier $n$, the optimal power $p_{kn} > 0$ only if $p_{{\bar k}
n} > 0$ for any $k \not= {\bar k}$.  Thus, on each subcarrier $n$,
allocating power to user $\bar k$ is preferred for optimality.
Furthermore, due to the uniform channel gain for user $\bar k$, the
optimal power allocation for user $\bar k$ is to allocate equal power
$\frac{1}{N}$ on every subcarrier.  Then the remaining problem is
equivalent to the OFDMA resource allocation problem in \cite{limit},
and the result follows.
\end{IEEEproof}

From the results in this section, using SR utility function instead of
WSR does not change the intractability of \OP{}.

\section{tractability analysis for relaxed \OP{}}

In this section, we aim to identify and characterize tractable cases
for \OP{}.  We provide tractability and convexity analysis for a
relaxed version of \WOP{} and \SOP{}.  For problem's relaxation, we
make the following observations.  First, from the proofs of Theorem
\ref{th:wnphard} and \ref{the:sop}, we have the following corollary.
\begin{corollary}
\label{the:noptot}
Both \SOP{} and \WOP{} remain NP-hard, even if constraint
(\ref{eq:p11}) is relaxed (i.e., the constraint is removed from the
optimization formulations $P1_{\text SR}$ and $P1_{\text WSR}$).
\end{corollary}  
Second, solving JPCAP will be challenging if (\ref{eq:p12}) is
present.  The same conclusion can be also applied to OFDMA resource
allocation, see e.g., \cite{ofdma, limit, yafeng}.  Next, the discrete
$x$-variables are introduced in \OP{} due to the presence of
constraint (\ref{eq:p13}).  This results in a non-convex feasible
region.  In the following, we relax two constraints (\ref{eq:p12}) and
(\ref{eq:p13}) as well as removing $x$-variables, and construct a
relaxed version of \SOP{} and \WOP{} in $P2_{\text SR}$ and $P2_{\text
WSR}$, respectively.
\begin{equation}
\label{p2sr}
P2_{\text SR}:  \  \  \  \
 \max_{\bm{p}} \limits  \ 
  \sum_{k \in \CK}\sum_{n \in \CN} R_{kn},  \ \ 
  \text{s.t.   \ \ (\ref{eq:p11}) } 
\end{equation}
\begin{equation}
\label{p2wsr}
P2_{\text WSR}:  \  \  \  \
 \max_{\bm{p}} \limits \ 
  \sum_{k \in \CK}w_k\sum_{n \in \CN} R_{kn},  \ \ 
  \text{s.t. \ \ (\ref{eq:p11})} 
\end{equation}
Note that both formulations above are with the $p$-variables only.  We
characterize the optimal power allocation for $P2_{\text{SR}}$ in
SC-NOMA first.

\begin{lemma}
\label{th:ofdma}
For $P2_{\text{SR}}$ in SC-NOMA with $g_1 \geq g_2 \geq ,\dots, \geq g_K$,
power allocation $p_1=P_{tot}, \ p_2=\dots =p_{K}=0$ is optimal.  
\end{lemma}  
\begin{IEEEproof}
First, observe that relaxing the user-individual power constraint
(\ref{eq:p12}) is equivalent to setting $P_k = P_{tot}$, i.e., the
user power limit is set to be equal to the total power limit, such
that (\ref{eq:p12}) becomes redundant.  Then, the result of the lemma
is obtained by applying the result of Theorem~\ref{th:noskip}, that
is, R-JPCAP is tractable and the optimum can be computed by using
Algorithm~\ref{alg:poly}.  Applying Algorithm~\ref{alg:poly} with $P_k
= P_{tot}$ leads immediately to the result of the lemma.
\end{IEEEproof}
Next, we generalize the results of Lemma \ref{th:ofdma} to
multi-carrier systems, and show $P2_{\text{SR}}$ is also tractable in
MC-NOMA.

\begin{lemma}
\label{th:ofdmamc}
For $P2_{\text{SR}}$ in MC-NOMA, there is an optimal solution
satisfying $|\CU_n| \leq 1$, $\forall n \in \CN$, i.e., OMA is
optimal.
\end{lemma}  
\begin{IEEEproof}
Suppose at the global optimum, a subcarrier $n$ has $|\CU_n| >
1$. Consider the two users having the largest gain values in $\CU_n$,
and, without any loss of generality of the proof, suppose that the
two user indices are 1 and 2, with $g_{1n} \geq g_{2n}$. Denote the power
allocated to users $1$ and $2$ on this subcarrier by $p_{1n}$ and
$p_{2n}$, respectively, with $p_{1n}>0$ and $p_{2n}>0$.  Denote by
$p^*_n$ the sum of the two, i.e., $p^*_n = p_{1n}+p_{2n}$. The
achieved sum utility for these two users is thus
$R_n^{(2)}=\log\frac{p_{1n}g_{1n}+\eta}{\eta}+\log\frac{p^*_ng_{2n}+\eta}{p_{1n}g_{2n}+\eta}$.

Consider allocating the amount of power $p^*_n$ to user 1 instead,
leaving zero power to user 2. The power allocation of the other users
remain unchanged.  Note that this power re-allocation affects only the
utility values of user 1 and 2. For user 1 the resulting utility
becomes $R_n^{(1)}=\log\frac{p^*_ng_{1n}+\eta}{\eta}$, whereas for
user 2 the utility is zero. Comparing ${R_n^{(1)}}$ and ${R_n^{(2)}}$,
we obtain the following.

\vspace{-2mm}
\begin{equation}
\label{eq:logrr}
\begin{aligned}
&{R_n^{(2)}}-{R_n^{(1)}}=
\log(\frac{p_{1n}g_{1n}+\eta}{p^*_ng_{1n}+\eta} \times \frac{p^*_ng_{2n}+\eta}{p_{1n}g_{2n}+\eta}) 
=\log(\frac{p_{1n}g_{1n}p^*_ng_{2n}+p_{1n}g_{1n}\eta+p^*_ng_{2n}\eta+\eta^2}{p_{1n}g_{1n}p^*_ng_{2n}+p^*_ng_{1n}\eta+p_{1n}g_{2n}\eta+\eta^2})
\end{aligned} 
\end{equation} 
It can be observed that
\begin{equation}
\begin{aligned}
&(p_{1n}g_{1n}\eta+p^*_ng_{2n}\eta)-(p^*_ng_{1n}\eta+p_{1n}g_{2n}\eta) =(p_{1n}\eta-p_n^*\eta)(g_{1n}-g_{2n}) < 0
\end{aligned} 
\end{equation}  
Since $p_{1n} < p_n^*$ and $g_{1n} > g_{2n}$, we have ${R_n^{(2)}} < {R_n^{(1)}}$. This contradicts the 
optimality of the first power allocation, and the lemma follows.
\end{IEEEproof}

From Lemma \ref{th:ofdmamc} for $P2_{\text{SR}}$, which amounts to
maximizing the sum rate utility subject to one single constraint on
the total power over all subcarriers, OMA resource allocation is
optimal. We remark that $P2_{\text{SR}}$ is convex and tractable.  It
can be checked that the Hessian matrix of the objective function in
$P2_{\text{SR}}$ is negative semi-definite.  As the $p$-variables are
continuous and (\ref{eq:p11}) is linear, $P2_{\text{SR}}$ is convex.
In general, the optimal solution can be obtained by performing a
polynomial-time algorithm \cite{yafeng}, that is, choosing the user
with the best channel gain on each subcarrier and then applying
water-filling power allocation for the assigned users.  The conclusion
is summarized below.

\begin{corollary}
The optimization problem in $P2_{\text{SR}}$ is tractable for SC-NOMA and MC-NOMA.
\end{corollary}  

\begin{remark}
In some proposed NOMA schemes, see e.g., \cite{system,nomainf}, the
users with inferior channel condition may request more power to
enhance user fairness, e.g., if $g_{1n} \geq g_{2n} \geq ,\dots, \geq
g_{Kn}$ on a subcarrier $n$, the power allocation is subject to
$0<p_{1n} \leq p_{2n} \leq, \dots, \leq p_{Kn}$.  By the results of
Lemma \ref{th:ofdma}, we remark that, on each subcarrier $n$, equal
power allocation for the multiplexed users $k \in \CU_n$ is
optimal. $\Box$
\end{remark}

In the following, we characterize the tractability and convexity for
$P2_{\text{WSR}}$.  From formulation $P2_{\text{WSR}}$, the convexity
is not straightforward to obtain.  Note that Lemma \ref{th:ofdma} and
Lemma \ref{th:ofdmamc} may not hold for $P2_{\text{WSR}}$, as a user
with inferior channel gain may be associated with higher weight, and
as a result, the optimum possibly has $|\CU_n| > 1$ for some $n \in
\CN$.  We make the following derivations to show
$P2_{\text{WSR}}$ is convex.

\begin{theorem}
The optimization problem in $P2_{\text{WSR}}$ is convex. 
\end{theorem}  
\begin{IEEEproof}
From Eq.~(\ref{eq:rkn}), for any subcarrier $n \in \CN$, there are $K$
equations linking the power allocation with the user rates.  For the
user with the best gain value, there is no interference term in
Eq.~(\ref{eq:rkn}), and hence the power variable of this user can be
expressed in its rate. Going through the remaining users in descending
order of gain and performing successive variable substitution, the
$p$-variables can be all expressed in the rate values.  Utilizing the
observation, we prove the convexity by reformulating $P2_{\text{WSR}}$
by treating rates $R_{kn}, k \in \CK, n
\in \CN$ as the optimization variables. This 
transformation is analogous to the geometric programming method
\cite{gp}. To facilitate the proof, we use $m_n(i)$ to denote the user index in
the $i$th position in the sorted sequence for subcarrier $n$, with
indices $i=0, \dots, K$, and the convention that $m_n(0)=0$.  
Problem $P2_{\text{WSR}}$ is then reformulated below.

\begin{subequations}
\label{reformulate}
\begin{align}
  \qquad & \hspace{-7mm} \max_{\bm{R}} \limits\ 
  \sum_{i=1}^{K} w_k \sum_{n=1}^{N} R_{kn} \qquad &  \\
\qquad &  \hspace{-7mm} \text{s.t.}
\sum_{i=1}^{K}\sum_{n=1}^{N} 
(\frac{\frac{\eta}{g_{m_n(i),n}}-\frac{\eta}{g_{m_n(i-1),n}}}{P_{tot}+\sum_{\bar n=1}^{N} \frac{\eta}{g_{m_{\bar n}(i-1),\bar n}} } ) 
\exp({\sum_{h=i}^{K} R_{m_n(h), n}}) \leq 0  \label{recon1} & \\ 
\qquad & \hspace{-3mm}  R_{kn} \geq 0, \ \forall k \in \CK, \ \forall n \in \CN \qquad &  \label{recon2}
\end{align}
\end{subequations}

The objective and constraints (\ref{recon2}) are both linear.
For constraints (\ref{recon1}), note that
$\frac{\eta}{g_{m_n(i),n}}-\frac{\eta}{g_{m_n(i-1),n}}\geq 0$, due to
the descending order of channel gains, Hence the $sum$-$exp$ function
in (\ref{recon1}) is convex
\cite{convex}, and the theorem follows.
\end{IEEEproof}

For the above convex and tractable cases, i.e., $P2_{\text{SR}}$ and
$P2_{\text{WSR}}$, standard optimization approaches for convex problem
can be applied.  For intractable cases, we develop an algorithmic
framework based on Lagrangian dual optimization and dynamic
programming (DP) to provide both near-optimal solutions and optimality
bounds in the next section.

\section{Optimization Algorithm for NOMA power and channel allocation}

In view of the complexity results, we aim to develop an algorithm that
is not for exact global optimum, yet the algorithm by design, is
capable of providing near-optimal solutions.  Moreover, the algorithm
is expected to deliver optimality bounds in order to gauge
performance, and is capable of progressively improving the bounds by
scaling parameters.  In this section, we propose an algorithmic
framework based on Lagrangian duality and dynamic programming
(\ALG{}).  In the developed algorithm, we make use of the Lagrangian
dual from relaxing the individual power constraint (\ref{eq:p12}) with
multipliers, and we develop a DP based approach to solve the problem
for given multipliers.  The algorithm is designed to solve both \SOP{}
and \WOP{} problems.  For generality, we take \WOP{} for illustration.

\subsection{Lagrangian Duality and Power Discretization}

Let vectors $\bm{p}$ and $\bm{x}$ collect all $p$-variables and
$x$-variables, respectively. Vector $\bm{\lambda} := \{\lambda_{k}, \
\forall k \in \CK\}$ contains the Lagrangian multipliers associated
with constraints (\ref{eq:p12}) in $P1_{\text{WSR}}$.  We construct
the subproblem of Lagrangian relaxation below.
\begin{subequations}
\begin{align}
P_{\text{LR}}: \notag 
\qquad & \hspace{-7mm}  \max_{\bm{x}, \bm{p}} \limits \hspace{-0mm} {L}(\bm{x}, \bm{p}, \bm{\lambda})\hspace{-1mm}=\hspace{-1mm}\sum_{k \in \CK}\hspace{-1mm}w_k\hspace{-1mm}\sum_{n \in \CN}\hspace{-1mm}x_{kn}R_{kn} \hspace{-1mm}+\hspace{-1mm} \sum_{k \in \CK}\hspace{-1mm}\lambda_k(P_k-\hspace{-1mm}\sum_{n \in \CN}\hspace{-1mm}p_{kn})  \notag  \\  
\qquad &   \hspace{-3mm} \text{s.t.}  \
\text{(\ref{eq:p11}), (\ref{eq:p13})} \notag &  
\end{align}
\end{subequations}
$P_{\text{LR}}$ is subject to the total power constraint
(\ref{eq:p11}) as well as constraints (\ref{eq:p13}) that limit the
number of users in each subcarrier.  Unlike the objective in
$P1_{\text{WSR}}$ and $P1_{\text{SR}}$, allocating power to user $k$
on subcarrier $n$ in $P_{\text{LR}}$ requires to pay a penalty in
utility, i.e., $-\lambda_kp_{kn}$.  The Lagrange dual function is
defined by $z(\bm{\lambda})=\max_{\bm{x}, \bm{p}} \limits {L}(\bm{x},
\bm{p}, \bm{\lambda})$. The dual optimum is correspondingly defined below.

\vspace{-2mm}
\begin{equation}
\label{eq:minG}
z^*=\min_{\bm{\lambda} \succeq 0}z(\bm{\lambda})
\end{equation}  

The optimization task amounts to solving $P_{\text{LR}}$ for a given
$\bm{\lambda}$ and finding the optimal $\bm{\lambda}$ to minimize the
Lagrangian dual in \eqref{eq:minG}.  Note that since $P1_{\text WSR}$
and $P1_{\text SR}$ are non-convex in general, there may exist a
duality gap between $z^*$ and global optimum $z^{\dagger}$ to the
original problem, i.e., $z^* \geq z^{\dagger}$.

Formulation $P_{\text{LR}}$ is non-convex in general due to the
reasons we discussed in Section \ref{sec:opf}. We consider solving
$z(\bm{\lambda})$, making use of the observation that, once the power
is discretized, $P_{\text{LR}}$ admits the use of DP for reaching 
optimality in polynomial-time
of the problem size and the number of power discretization levels. To
this end, we discretize the power budget $P_{tot}$ into $J$ uniform
steps, and denote by $\delta$ the size of each step, i.e., $\delta = P_{tot} / J$.
Denote by $p^j$ be the power value for level $j$ 
and $p^j = \delta * j$, where $j
\in \CJ=\{1, \dots, J\}$.  We denote by $P_{\text{LR-D}}$ the version of $P_{\text{LR}}$ after
power discretization.  The formulation of $P_{\text{LR-D}}$ and its
optimization variables are presented below.

\vspace{3mm}
$\hspace{-3mm} x_{kn}^j$ = 
    $\left\{
      \begin{array}{l l}      
      \hspace{-1mm}  1 & $if power level $j$ is allocated to user $k$ on $  $subcarrier $n$, $ \\
     \hspace{-1mm}   0 & $otherwise.$ \\
      \end{array}
    \right.$

\begin{subequations}
\label{eq:plr}
\begin{align}
\qquad &  \hspace{-1.2cm} P_{\text{LR-D}}: \hspace{3mm} \max_{\bm{x}} \limits \hspace{-0mm}{L_D}(\bm{x}, \bm{\lambda})\hspace{-1mm}= \hspace{-1mm}\sum_{k \in \CK}\hspace{-1mm}w_k \hspace{-1mm} \sum_{n \in \CN}\sum_{j \in \CJ} x_{kn}^jR_{kn}^j \hspace{-1mm}+  \sum_{k \in \CK}\hspace{-1mm}\lambda_k(P_k-\hspace{-2mm}\sum_{n \in \CN}\sum_{j \in \CJ} x_{kn}^jp^j) \label{eq:plrobj}
 \qquad &  \\
\label{eq:plr1}
\qquad &  \hspace{-2mm}  \text{s.t.}  \ \ \ 
   \sum_{k \in \CK}\sum_{n \in \CN}\sum_{j \in \CJ} x_{kn}^jp^j  \leq P_{tot} &\\
\label{eq:plr2}
\qquad & \hspace{7mm}  \sum_{k \in \CK}\sum_{j \in \CJ} x_{kn}^j \leq M,  \ \forall n \in \CN & \\
\label{eq:plr3}
\qquad &  \hspace{7mm}     \sum_{j \in \CJ} x_{kn}^j  \leq 1,  ~\forall k \in \CK, ~\forall n \in \CN 
\end{align}
\end{subequations}

The objective (\ref{eq:plrobj}) and constraints (\ref{eq:plr1}),
(\ref{eq:plr2}) originate from $P_{\text{LR}}$ but are adapted to
power discretization.  In (\ref{eq:plrobj}), the achievable rate of
allocating user $k$ on subcarrier $n$ with power level $j$ is denoted
by $R_{kn}^j$ below for $\forall k \in \CK, \ \forall n \in \CN$,
$\forall j \in \CJ$, and with normalized bandwidth $\frac{B}{N}=1$.
\begin{equation}
R_{kn}^j=\log(1+\frac{p^jg_{kn}} { \sum_{\substack{h \in \CK \backslash \{k\}: \\  b_n(h)<b_n(k)}}   \limits (\sum_{\substack{j' \in \CJ}} \limits x_{hn}^{j'}p^{j'}) g_{kn} +\eta})
\notag 
\end{equation}
Constraints \eqref{eq:plr3} state that each user on a subcarrier can
select one power level at most, and $\sum_{j \in \CJ} x_{kn}^j = 0$
means that there is no power allocation for user $k$ on subcarrier
$n$.  For given $\bm{\lambda}$, let
$z_D(\bm{\lambda})=\max_{\substack{\bm{x}}} \limits \hspace{-0mm}
{L_D}(\bm{x}, \bm{\lambda})$ and ${\bm p}^*$ denote the optimal
objective value and the corresponding power solution of
$P_{\text{LR-D}}$, respectively.  Next, we develop a DP based approach
to solve $P_{\text{LR-D}}$ exactly to optimality.

\begin{remark}
Power discretization in $P_{\text{LR-D}}$ is considered as an
approximation for the continuous power allocation in $P_{\text{LR}}$.
However, in practical systems, the power is typically set in discrete
steps, e.g., discrete power control in LTE downlink \cite{r10}.  In
this case the discrete model $P_{\text{LR-D}}$ is exact.  $\Box$
\end{remark}

\subsection{Two-Stage DP Based Approach}

Given power levels in set $\CJ$ and multipliers in vector
$\bm{\lambda}$, problem $P_{\text{LR-D}}$ can be solved by using DP.  In
general, DP guarantees global optimality if the problem has the so
called ``optimal substructure property'' \cite{networkflow}.  A
classical example is the knapsack problem with integer coefficients
\cite{KePfPi04}.  In our case, $P_{\text{LR-D}}$ does exhibit the
property, and a proof of the optimality of DP will be provided later
in Theorem~\ref{dp}.

To ease the presentation, we describe the DP algorithm in two stages.
In the first stage, intra-subcarrier power
allocation is carried out among users, that is, for subcarrier $n \in
\CN$, the algorithm computes the optimal user power allocation by
treating power $p^j$, $j=1, \dots, J$, as the power budget for the
subcarrier in question.  The optimal utility value of consuming power
$p^j$ on $n$ is denoted by $V_{n, j}$, where $n \in \CN, \ j \in \CJ$.
Since the number of multiplexed users cannot exceed $M$ on each
subcarrier, we keep track on the optimum allocation for each $m \in
\{1, \dots, M\}$.  We define a tuple of format $\Ct=(u_{\Ct},
M_{\Ct})$ to represent a candidate partial solution, where $u_{\Ct}$
is the utility value, and $M_{\Ct}$ is the number of users allocated
with positive power. For a partial problem of assigning positive
power to exactly $m$ out of the first $k$ users with power budget
$p^j$, the optimal utility is denoted by $T_{k,j}^m$.

The optimality of stage 1 is obtained from DP recursion.  The values
$T_{k,j}^m$ can be arranged in form of a $K \times J \times M$ matrix
$\mathcal{A}_1$.  Computing $T_{k,j}^m$ for $k, m, j=1$ is
straightforward.  For $k\geq 2, m\geq 2$ and $j \geq 2$, the following
recursive formula is used to obtain the corresponding value in
$\mathcal{A}_1$.
\vspace{-1mm}
\begin{equation}
\label{eq:dp1}
T_{k,j}^m=\max\{\max_{\substack{j'=1,\dots,j-1}}\{w_kR_{kn}^{j'}-\lambda_kp^{j'}+T_{k-1,j-j'}^{m-1}\}, \ T_{k-1,j}^{m}\}
\end{equation} 
From (\ref{eq:dp1}), the procedure of obtaining $T_{k,j}^m$ is
decomposed into multiple stages, and the recursion is applied to move
from one stage to another.  Thus, each partial problem has an optimal
substructure \cite{networkflow}.

The algorithmic operations for stage one are given in Algorithm
\ref{alg:single}.  The bulk of the computation starts at Line
\ref{line:k1}.  For user $k=1$, exactly one tuple is created for each
$j$, and the corresponding utility value is $T_{1,j}^1, \ j \in \CJ$,
see Line \ref{line:k1t}.  For users $k>1$, the recursion is performed
in Lines \ref{line:dpk2} to \ref{line:ot}. In Lines \ref{line:new0} to
\ref{line:ot}, a tuple $\Ct$ is created, and its utility value
$u_{\Ct}$ will replace the current $T_{k,j}^m$ if $u_{\Ct}>T_{k,j}^m$.
Note that in Line \ref{line:createt}, the utility $u_{\Ct}$ is the sum
of two parts, i.e., assigning a trial power $p^{j'}$ to user $k$ plus
the previously obtained maximum utility $T_{k-1,j-j'}^{m-1}$.  The
algorithm terminates when all the $K$ users and $J$ power levels have
been processed.  The optimal value of assigning power $p^j$ on a
subcarrier $n$ is stored in $V_{n,j}$, see Line \ref{line:vnl}.

\begin{algorithm}[ht!]  
{\bf Input}: $K$, $J$, $M$, and $\bm{\lambda}$ \\
{\bf Output}: $V_{n, j}$ for each $j \in \CJ$ on subcarrier $n$
\vspace{-1mm}
\caption{Stage 1 of \DP{}: Intra-subcarrier Power Allocation}
\label{alg:single}
\begin{algorithmic}[1]
\STATE $T_{k,j}^m \gets \emptyset$,  for $\forall k \in \CK, \forall j \in \CJ, \forall m \in \{1, \dots, M\}$
\FOR{$j=1:J$}   \label{line:k1}
\STATE $\Ct \gets (w_1R_{1n}^{j}-\lambda_1p^j$, $1)$
\STATE $T_{1,j}^1 \gets u_\Ct$ 
\label{line:k1t}
\ENDFOR
\FOR{$k=2:K$} \label{line:dpk2}
   \FOR{$j=1:J$} 
  \FOR{$j'=0:j$} 
  \FOR{$m=1:\min \{k, M\}$} 
\IF{$j'=0$}  \label{line:new0}
\STATE $\Ct \gets (T_{k-1,j}^m$, $m$), $T_{k,j}^m \gets  \max \{u_{\Ct},  T_{k,j}^m\}$  
\ENDIF
\IF{$1 \leq j' \leq j-1$ and $m < \min \{k, M\}$} 
\STATE $\Ct \gets (w_kR_{kn}^{j'}-\lambda_kp^{j'}+T_{k-1,j-j'}^m$, $m+1$)  \label{line:createt}
\STATE $T_{k,j}^{m+1} \gets  \max \{u_{\Ct},  T_{k,j}^{m+1}\}$  
\ENDIF
\ENDFOR   
\IF{$j'=j$} 
\STATE $\Ct \gets (w_kR_{kn}^{j}\hspace{-2mm}-\hspace{-1mm}\lambda_kp^{j}$, 1),    $T_{k,j}^1 \hspace{-1mm}\gets \hspace{-1mm}  \max \{u_{\Ct},  T_{k,j}^1\}$   \label{line:ot} 
\ENDIF
\ENDFOR 
\ENDFOR 
\ENDFOR
\FOR{$j=1:J$}  
\STATE {$V_{n,j} \gets \displaystyle \max_{k \in \CK,  m \in
\{1,\dots,M\}} T_{k,j}^m$} \label{line:vnl} 
\ENDFOR
\end{algorithmic}
\end{algorithm}

In the second stage, power allocation of $P_{tot}$ is carried out
among the subcarriers, i.e., inter-subcarrier power allocation.  Then
DP is applied to perform optimal power allocation at the subcarrier
level.  For given $\bm{\lambda}$, $z_D(\bm{\lambda})$ is obtained in
Algorithm \ref{alg:twostage}.  The operations start at Line
\ref{line2:alg2}.  In the first stage of \DP{}, Algorithm
\ref{alg:single} is performed to obtain $V_{n,1}, \dots, V_{n, J}$ for
each subcarrier $n \in \CN$.  Note that by the construction of
Algorithm \ref{alg:single}, in $V_{n,j}$, the number of multiplexed
users is at most $M$ for $\forall n \in \CN$. Thus in stage two, index
$m$ is no longer needed.  In Lines \ref{line2:dpb}-\ref{line2:dpe},
based on the accumulated value $\hat T_{n-1,j-j'}$, a new candidate
for $\hat T_{n,j}$ is obtained by adding $V_{n,j'}$.  Then for the
partial problem of allocating power $p^j$ to $1, \dots, n$
subcarriers, the optimal solution $\hat T_{n,j}$ is obtained.

\begin{algorithm}[ht!]
{\bf Input}: $K$, $N$, $J$, $M$, and $\bm{\lambda}$ \\
{\bf Output}: $z(\bm{\lambda})_D$
\caption{Two-Stage Dynamic Programming (\DP{})}
\label{alg:twostage}
\begin{algorithmic}[1]
\STATE Initialize $\hat T_{n,j} \gets 0$ for $\forall n \in \CN, \forall j \in \{0, \dots, J\}$
\STATE {\bf Stage 1:}
\FOR{$n=1:N$}    \label{line2:alg2}
\STATE Perform Algorithm \ref{alg:single}, and obtain $V_{n,1}, \dots, V_{n,J}$ 
\ENDFOR 
\STATE  {\bf Stage 2:}
\FOR{$j=1:J$}    \label{line2:dpb}
\STATE $\hat T_{1,j} \gets V_{1,j}$ 
\ENDFOR 
\FOR{$n=2:N$}   
\FOR{$j=1:J$} 
\STATE $\hat T_{n,j}= \displaystyle \max_{\substack{j'=1,\dots,j}}
\{(V_{n,j'}+\hat T_{n-1,j-j'}), \hat T_{n-1,j} \}$  \label{line2:dpe}
\ENDFOR 
\ENDFOR 
\STATE $z_D(\bm{\lambda}) \gets \displaystyle \max_{n \in \CN, j \in \CJ} \ \hat T_{n,j}$   \label{line2:g}
\end{algorithmic}
\end{algorithm}

The DP recursion (for $n \geq 2$) for the second stage is given in
Line \ref{line2:dpe}.  The values $\hat T_{n,j}$ for all $k \in \CK, \
j \in \CJ$ can be viewed in form of an $N \times J$ matrix
$\mathcal{A}_2$.  From the DP recursions in \DP{}, the global optimum
of $P_{\text{LR-D}}$ is obtained from accumulating the solutions of
the partial problems. By the end of stage two, $z_D(\bm{\lambda})$ is
equal to the maximum $\hat T_{n,j}$ among the elements in
$\mathcal{A}_2$, see Line \ref{line2:g}.  In $P_{\text{LR-D}}$, for
the partial problem for users $1,\dots,k$, subcarriers $1,\dots,n$,
and with a total power budget $p^j$, the optimum is independent of
that for the remaining subcarriers or users. The complexity for
computing optimality is provided in Theorem~\ref{dp}.

\begin{theorem}
\label{dp}
The global optimum of $P_{\text{LR-D}}$ is obtained by \DP{} with a
time complexity being polynomial in $M, N, K$, and $J$.
\end{theorem}
\begin{IEEEproof}
The input of $P_{\text{LR-D}}$ are $N$, $K$, $M$ and $J$.  By
inspecting \eqref{eq:dp1} and Line \ref{line2:dpe} in Algorithm
\ref{alg:twostage}, computing matrix $\mathcal{A}_1$ for all
subcarriers requires $\mathcal O(KNMJ^2)$ in running time.  For matrix
$\mathcal{A}_2$, the running time is of $\mathcal O(NJ^2)$.  Hence the
former is dominating, and the overall time complexity is $\mathcal
O(KNMJ^2)$, which is polynomial in the input size. \end{IEEEproof}

\begin{remark}
Increasing $J$ provides better granularity in
power discretization. By improving the granularity,
the solution of $P_{\text{LR-D}}$ can approach
arbitrarily close to that of $P_{\text{LR}}$. $\Box$
\end{remark}

\subsection{Algorithmic Framework: Lagrangian Duality With Dynamic Programming}

We develop a framework \ALG{} to deliver near-optimal solutions
(N-\ALG{}) of the global optimum $z^{\dagger}$ of $P1_{\text WSR}$ and
$P1_{\text SR}$.  We also derive a scheme UB-\ALG{} to provide upper
bounds for gauging the solution quality of N-\ALG{}.
\ALG{} is summarized in Algorithm \ref{alg:lddp}.  
In Line \ref{line:while} to Line \ref{line4:gd}, we obtain
$z_D(\bm{\lambda})$ and the power solution ${\bm p}^*$ by applying
\DP{} to solve $P_{\text{LR-D}}$.  These steps constitute N-\ALG{}.
The iterations for solving the Lagrangian dual terminate either after
a specified number of iterations $C_{max}$, or if the difference
between the objective values in two successive iterations is less than
$\epsilon$ \cite{networkflow}. Lines \ref{line4:ptot}--\ref{line4:overcomp}
form UB-\ALG{}, which approximates the global optimum from above.
UB-\ALG{} delivers an upper bound, i.e., a value that is guaranteed to
be no smaller than the global optimum. The purpose is for performance
evaluation. Note that the problem is maximization, and hence the
solution from the first part of the algorithm, N-\ALG{}, has a utility
value that is lower than global optimum. 
Computing the global optimum is NP-hard. However,
to assess the performance of
this solution, we can instead obtain the upper bound at significantly lower complexity.  
The deviation from global optimum cannot be more than the deviation from the upper bound, which can be used in our performance evaluation.

\begin{algorithm}[ht!]
\caption{\ALG{} for Solving \WOP{} }
\label{alg:lddp}
\begin{algorithmic}[1]
\STATE Initialize $\bm{\lambda}$, tolerance $\epsilon$, number of iteration $C \gets  0$, maximum number of iterations $C_{max}$, $\check d$ and $\hat d$ such that $|\check d - \hat d|  > \epsilon$, $\bm{p^*} \gets \bm{0}$
\WHILE {$|\check d - \hat d|  > \epsilon$ or $C \leq C_{max}$}  
\STATE $\check d \gets \hat d$ \label{line:while}
\STATE  Perform \DP{} (Algorithm \ref{alg:twostage}) to solve $P_{\text{LR-D}}$
\STATE $z_D(\bm{\lambda}) \gets \displaystyle \max_{n \in \CN, j \in \CJ} \hat T_{n,j}$
\STATE ${\bm p}^*$ $\gets$ power solution of $z_D(\bm{\lambda})$
\STATE $\hat d \gets z_D(\bm{\lambda})$
\IF{${\bm p}^*$ violates constraints \eqref{eq:p12}} 
\STATE {Convert ${\bm p}^*$ to ${\bm p}_f$}
\label{line4:convert}
\ENDIF
\STATE Compute ${f}({\bm p}_f)_{\text W}=\sum_{k \in \CK}w_{k}\sum_{n \in \CN}R_{kn}$ by (\ref{eq:rkn})  \label{line4:lb}
\STATE $V_{\text{LB}} \gets {f}({\bm p}_f)_{\text W}$  \label{line4:gd}
\STATE Update $\bm{\lambda}$ by subgradient method  \label{line4:sub}
\STATE $C=C+1$
\ENDWHILE 
\STATE {\bf end}
\STATE Relax (\ref{eq:plr1}) with $\mu$ and construct $P_{\text{LR-D}}'$   \label{line4:ptot}
\REPEAT \label{line4:bisec}
\STATE Bisection search for $\mu$
\STATE  $z_D(\bm{\lambda}, \mu) \gets \max_{\bm{x}} \limits {L_D'}(\bm{x}, \bm{\lambda}, \mu)$
\UNTIL{$\mu^* \gets \{\mu: \min_{\mu \geq 0} \limits \max_{\substack{\bm{x}}} \limits {L_D'}(\bm{x}, \bm{\lambda}, \mu) \}$ } \label{line4:bisecend}
\STATE $V_{\text{UB}} \gets z_D(\bm{\lambda}, \mu^*)$
 \label{line4:overcomp}
\STATE {\bf Return}: $V_{\text{LB}}$ and $V_{\text{UB}}$
\end{algorithmic}
\end{algorithm}

In Line \ref{line4:convert}, the users' individual power constraints
may be violated in ${\bm p}^*$.  Thus, we develop a three-step
approach to convert ${\bm p}^*$ into a feasible solution ${\bm p}_f$
for \OP{} if the former violates \eqref{eq:p12}.  Let set $\bar \CK$
denote the users allocated with positive power in ${\bm p}^*$, $\bar
\CK \subseteq \CK$.  We denote $\CK' = \{k \in \bar \CK: \sum_{n \in
\CN} p_{kn} > P_k\}$ as the subset of users for which \eqref{eq:p12}
does not hold.  To obtain ${\bm p}_{f}$, step one, power allocation
for each user $k \in \bar \CK \setminus \CK'$ on subcarriers keeps
same as in $\bm p^*$.  Step two, for each $k \in \CK'$, the
subcarriers are sorted in ascending order in power allocation of $k$.
Following the sequence, power is allocated as in $\bm p^*$, however
until the limit $P_k$ is reached.  Doing so releases an amount of
$\sum_{k \in \CK'}\sum_{n \in \CN} p_{kn} - \sum_{k \in \CK'} P_k$
power that can be re-allocated to users in $\bar \CK \setminus \CK'$
in step three.  The re-allocation follows the descending order of the
product of channel gain and weight.  That is, letting $({\tilde k},
{\tilde n}) = \hbox{argmax}~ w_{kn}g_{kn}, k \in \bar \CK \setminus
\CK', n \in \CN$, $p_{{\tilde k}{\tilde n}}$ is increased as much as
allowed by $P_{\tilde k}$ and $P_{tot}$, then we select the next best
candidate, and so on, until either $P_{tot}$ or $P_k$ for all $k \in
\bar \CK$ is reached.  In Line \ref{line4:lb}, we obtain the resulting
utility ${f}({\bm p}_f)_{\text W}$ for using ${\bm p}_f$, where the
calculation of ${f}({\bm p}_f)_{\text W}=\sum_{k \in \CK}w_{k}\sum_{n
\in \CN}R_{kn}$ follows the equation in (\ref{eq:rkn}).  The utility
value of N-\ALG{} is delivered in $V_{\text{LB}}$ at Line
\ref{line4:gd}.
In Line \ref{line4:sub},
$V_{\text{LB}}$ is used in the calculation of step size of
subgradient optimization (see \cite{networkflow}).

To provide an upper bound for $z^{\dagger}$, we apply post-processing
to $z_D(\bm{\lambda})$ in \ALG{}.  We remark that if $J$ is
sufficiently large, $z_D(\bm{\lambda})$ can be empirically considered
as an upper bound to \WOP{} or \SOP{} due to Lagrangian duality,
however, theoretically there is no guarantee. For example,
$z_D(\bm{\lambda})$ could be possibly less than $z^{\dagger}$ for
small $J$, e.g., $J=1$.  From Line \ref{line4:ptot} to Line
\ref{line4:overcomp}, we design an approach to convert
$z_D(\bm{\lambda})$ to a theoretically guaranteed upper bound.  First,
in Line \ref{line4:ptot}, for the multipliers in $ \bm{\lambda}$ and
keeping their values fixed, we further relax the total power constraint
(\ref{eq:plr1}) with multiplier $\mu$, and reconstruct the subproblem
of Lagrangian relaxation below.
\begin{subequations}
\label{plambmu}
\begin{align}
P_{\text{LR-D}}':
\qquad & \hspace{-3mm} \max_{\bm{x}} \limits  {L_D'}(\bm{x}, \bm{\lambda}, \mu)=
\underbrace{   
\sum_{k \in \CK}w_k \hspace{-1mm} \sum_{n \in \CN}\sum_{j \in \CJ} x_{kn}^j \bar R_{kn}^j
}_{\text{part I}} 
+  \notag  \\  
\qquad &   \hspace{-1.8cm} 
\underbrace{   
\sum_{k \in \CK}\hspace{-1mm} \lambda_k(P_k\hspace{-1mm}-\hspace{-2mm}\sum_{n \in \CN}\hspace{-0.7mm}\sum_{j \in \CJ}\hspace{-0.7mm} x_{kn}^j(p^j\hspace{-1mm}-\hspace{-1mm}\delta))\hspace{-1mm}
}_{\text{part II}} 
+\hspace{-1mm}
\underbrace{ 
 \mu(P_{tot}\hspace{-1mm}-\hspace{-2mm}\sum_{\hspace{-0.7mm}k \in \CK}\hspace{-0.7mm}\sum_{n \in \CN}\hspace{-0.7mm}\sum_{j \in \CJ} \hspace{-0.7mm} x_{kn}^j(p^j\hspace{-1mm}-\hspace{-1mm}\delta)) 
}_{\text{part III}} 
 \notag \\
\qquad &   \hspace{-2mm} \text{s.t.}  \
\text{(\ref{eq:plr2}) and (\ref{eq:plr3})} \notag &  
\end{align}
\end{subequations}
The calculation of $\bar R_{kn}^j$ for $\forall k \in \CK, \forall n \in \CN$, and $\forall j \in \CJ$ is shown below.
\begin{equation}
\label{eq:overcal}
\bar R_{kn}^j=\log(1+\frac{(p^j+\delta)g_{kn}} {\hspace{-6mm} \sum_{\substack{h \in \CK \backslash \{k\}: \\ b_n(h)<b_n(k)}} \limits  \hspace{-1mm} (\sum_{\substack{j' \in \CJ}} \limits x_{hn}^{j'}p^{j'}-\delta x_{hn}^{j'}) g_{kn} +\eta})  
\end{equation}
Recall that $\delta$ is the step in power discretization.  In
comparison to ${L}(\bm{x}, \bm{p}, \bm{\lambda})$ in $P_{\text{LR}}$
and ${L_D}(\bm{x}, \bm{\lambda})$ in $P_{\text{LR-D}}$, the
construction of $P_{\text{LR-D}}'$ contains addition or subtraction of
power $\delta$, for the purpose of ensuring that the outcome is a valid
upper bound to the global optimum. This is achieved by using $\delta$
to obtain more optimistic values in all three parts of the
function. In \eqref{eq:overcal}, for example, one power step $\delta$
is added to the signal of interest in the numerator and each
interfering signal becomes weaker due to the subtraction of $\delta$
in the denominator. As a result, the overall utility for
${L_D'}(\bm{x}, \bm{\lambda},
\mu)$ is guaranteed to be an over-estimation.
From Lines \ref{line4:bisec} to \ref{line4:bisecend}, bisection search
is applied to obtain the optimal $\mu^*$ such that
$z_D(\bm{\lambda}, \mu^*)=\min_{\mu \geq 0} \limits
\max_{\substack{\bm {x}}} \limits {L_D'}(\bm{x}, \bm{\lambda}, \mu)$.
Then, the upper bound is delivered in $V_{\text{UB}}$ in Line
\ref{line4:overcomp}.  The validity of this upper bound is proved in
Theorem \ref{realub}.  

\begin{theorem}
\label{realub}
$V_{\text{UB}} \geq z^{\dagger}$.
\end{theorem}
\begin{IEEEproof}
Suppose $\bm{\lambda}^*$ is the multiplier vector for
$P_{\text{LR-D}}$ when Algorithm \ref{alg:lddp} terminates.  Note that
$\bm{\lambda}^*$ for $z(\bm{\lambda}^*)$ may not necessarily lead to
the minimum dual value $z^*$ in $P_{\text{LR}}$, so we have
$z(\bm{\lambda}^*) \geq z^* \geq z^{\dagger}$.  We prove that
$z_D(\bm{\lambda}^*, \mu^*) \geq z(\bm{\lambda}^*)$ holds, to show
$V_{\text{UB}}$ is an upper bound of $z^{\dagger}$.  We use vector
$\bm p_c \succ 0$ to denote the optimal power allocation for
$z(\bm{\lambda}^*)$.  Based on $\bm p_c$, we now construct a power
vector $\bm p_d \succ 0$ for $P_{\text{LR-D}}'$.  Each power value in
$\bm p_c$ is rounded to $\bm p_d$ such that each element in $\bm p_d$
is represented by the closest power level $j \in \CJ$, i.e., $\bm p_c
+\bm \theta=\bm p_d$, where $|\bm \theta| \preceq \delta$.  Given $\bm
p_c$ and $\bm p_d$, the corresponding $x$-vectors $\bm x_c$ and $\bm
x_d$ are derived for $P_{\text{LR}}$ and $P_{\text{LR-D}}'$,
respectively.  Substituting $\bm{\lambda}^*, \mu$ and $\bm x_d$ in the
objective of $P_{\text{LR-D}}'$, we have ${L_D'}(\bm x_d,
\bm{\lambda}^*, \mu)$.  Since we over-calculate the objective in
$P_{\text{LR-D}}'$, the summation of part I and
part II in ${L_D'}(\bm x_d, \bm{\lambda}^*, \mu)$ is greater than or
equal to ${L}(\bm x_c, \bm p_c, \bm{\lambda}^*)$.  Part III is no
less than zero in ${L_D'}(\bm x_d, \bm{\lambda}^*, \mu)$ for any $\mu
\geq 0$ due to $||\bm p_c||_1 \leq P_{tot}$ and $|\bm \theta| \preceq
\delta$.  Then ${L_D'}(\bm x_d, \bm{\lambda}^*, \mu) \geq {L}(\bm x_c,
\bm p_c, \bm{\lambda}^*)=z(\bm{\lambda}^*)$ for any $\mu \geq 0$.
Thus, we have ${L_D'}(\bm x_d, \bm{\lambda}^*, \mu^*) \geq
z(\bm{\lambda}^*)$.  Note that for $\lambda^*$ and $\mu^*$,
$V_\text{UB}=z_D(\bm{\lambda}^*, \mu^*)$ is the optimum value of the
Lagrangian relaxation in $P_{\text{LR-D}}'$, whereas ${L_D'}(\bm x_d,
\bm{\lambda}^*, \mu^*)$ is not because $\bm {x_d}$ is not necessarily
an optimal power allocation.  Therefore, $z_D(\bm{\lambda}^*, \mu^*)
\geq {L_D'}(\bm x_d, \bm{\lambda}^*, \mu^*)$, and the conclusion
follows.
\end{IEEEproof}

On the complexity of \ALG{}, we observe the following.  Within each
iteration, N-\ALG{} calls Algorithm~\DP{} that has polynomial-time
complexity $\mathcal O(KNMJ^2)$ and hence is scalable. Note that the
number of power levels $J$ can be tuned from the complexity
perspective. An iteration of N-\ALG{} may require the conversion to
feasible power allocation (Line~\ref{line4:convert}).  It is easily
observed that this three-step conversion, as outlined earlier, has a
complexity of $\mathcal O(KN \log_2(KN))$, which scales much better than $\mathcal O(KNMJ^2)$.  Next, obtaining the upper bound in
UB-\ALG{} consists of using Algorithm~\DP{} in one-dimensional
bi-section search. This computation does not lead to the computational
bottleneck, because the upper bound is computed only once, and its
purpose is not for power allocation in NOMA, but for
performance evaluation as a post-processing step. Hence, overall, the
complexity is determined by N-\ALG{}, and equals $\mathcal
O(CKNMJ^2)$, where $C$ is the number of subgradient optimization iterations upon
termination.  Subgradient optimization for Lagrangian duality has
asymptotic convergence in general. In the next section, however, we
observe that convergence is approached with only a few iterations.

\section{Performance Evaluation}
\label{sec:performance}

\subsection{Experimental Setup}

We have carried out performance studies in downlink with randomly and
uniformly distributed users.  Table \ref{tab:parameter} summarizes the
key parameters.  We generate one hundred instances and consider the
average performance.
To evaluate the performance of \ALG{}, we have implemented a previous
NOMA power and channel allocation scheme called ``fractional transmit
power control" (NOMA-FTPC) and an OFDMA scheme with FTPC (OFDMA-FTPC)
\cite{system}.  In these two schemes, the set of multiplexed users
$\CU_n$ for each subcarrier $n$ is determined by a greed-based user
grouping strategy, where $|\CU_n|=M$ for NOMA-FTPC and $|\CU_n|=1$ for
OFDMA-FTPC.  Based on the user allocation, the FTPC method is then
used for power allocation.  In FTPC, more power is allocated to the
users with inferior channel condition for the fairness consideration
\cite{system}.

\begin{table}[htbp]
\caption{Simulation Parameters.}
\vspace{-5mm}
\label{tab:parameter}
\centering
\begin{tabular}[t]{l l}
\hline
\bf{Parameter}  &   \bf{Value}   \\
Cell radius  & 200~m \\
Carrier frequency  & 2~GHz \\
Total bandwidth ($B$) & 4.5~MHz \\
Number of subcarriers ($N$) & 5 in NOMA, 25 in OFDMA \\
Number of users ($K$) & 4 to 20 \\
Path loss & COST-231-HATA\\
Shadowing & Log-normal, 8~dB standard deviation \\
Fading & Rayleigh flat fading \cite{tse}\\
Noise power spectral density & -173~dBm/Hz \\
Total power ($P_{tot}$) &  1~W   \\
Number of power levels ($J$) & 20 to 100 \\
Minimum power unit ($\delta$=$\frac{P_{tot}}{J}$) & 0.01 to 0.05~W \\
User power limit  ($P_k$) &  0.2~W   \\
Parameter $M$  & 2 to 6\\
Tolerance $\epsilon$ in \ALG{}  & $10^{-5}$ \\
$C_{\text{max}}$ in \ALG{}  & 200 \\
\hline
\end{tabular}
\end{table}

For OFDMA-FTPC, following the LTE standard, the overall bandwidth of
$4.5$ MHz is divided into twenty-five subchannels with the bandwidth
of $180$ kHz for each.  For NOMA implementation, considering the fact
that the decoding complexity and signaling overhead increase with the
number of subcarriers \cite{higuchi}, and following the NOMA setup in
\cite{system}, we consider five subcarriers with the bandwidth of
$900$ kHz for each in NOMA-FTPC and in the proposed \ALG{}.

In the simulations, N-\ALG{} aims to deliver a near-optimal solution
(also a lower bound), and UB-\ALG{}, by design, provides an upper
bound for global optimum.  In the following, we examine five
performance aspects.  First, a comparative study for the SR utility of
\ALG{}, NOMA-FTPC and OFDMA-FTPC is carried out.  Second, we evaluate the convergence behavior of \ALG{}.
Third, we consider
the WSR utility and examine users' fairness.  
Fourth, we evaluate the
throughput performance for the cell-edge users.
Fifth, we investigate the characteristics of user grouping in N-\ALG{}.

\subsection{Performance in Throughput and Bounding}

\begin{figure}[htbp]
\begin{minipage}[t]{0.45\textwidth}
\centering
\includegraphics[scale=0.36]{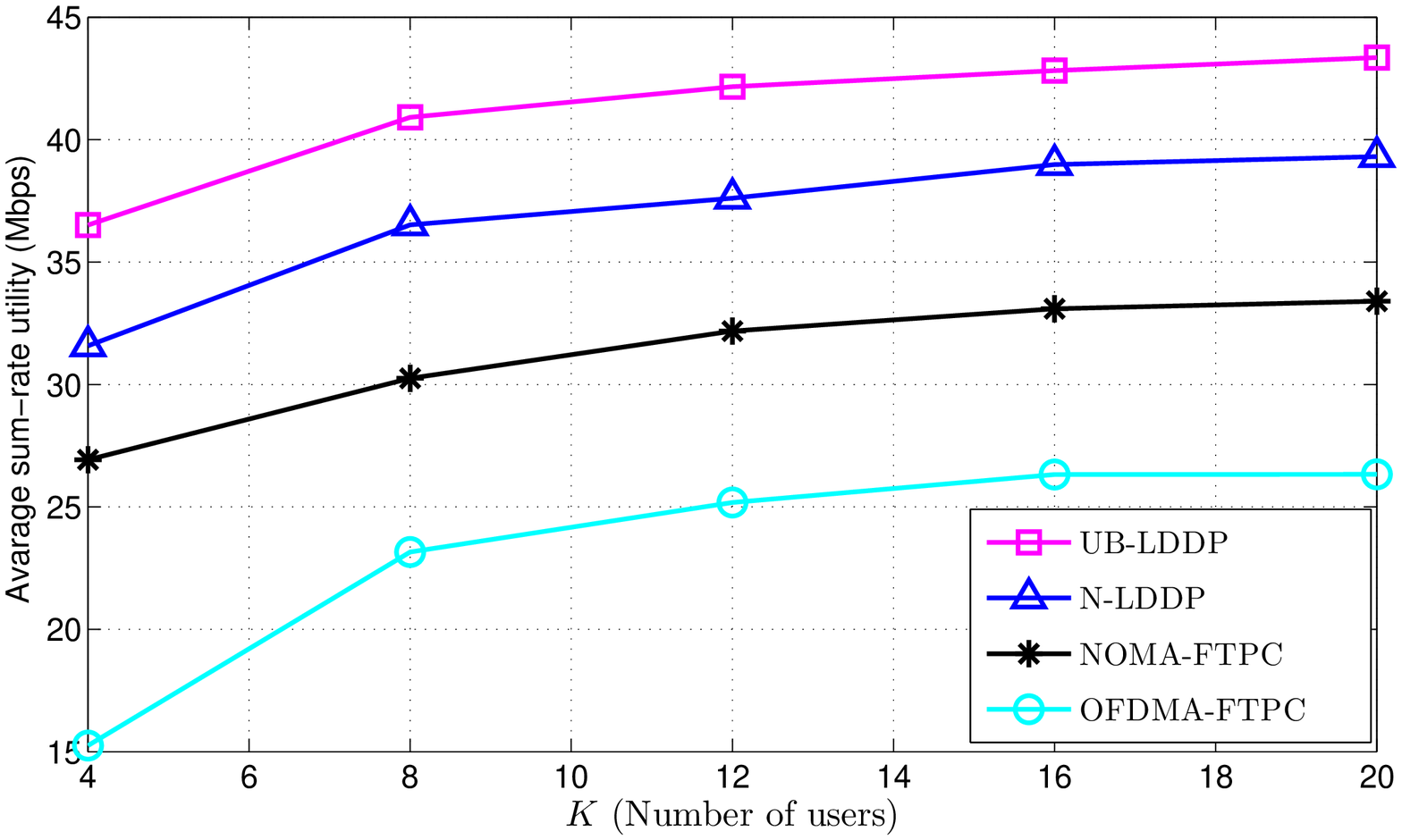}
\vspace{-1.3cm}
\caption{\ALG{} performance in respect of the number of users.}
\label{fig:SRuser}
\end{minipage}
\hspace{1.0cm}
\begin{minipage}[t]{0.45\textwidth}
\centering
\includegraphics[scale=0.37]{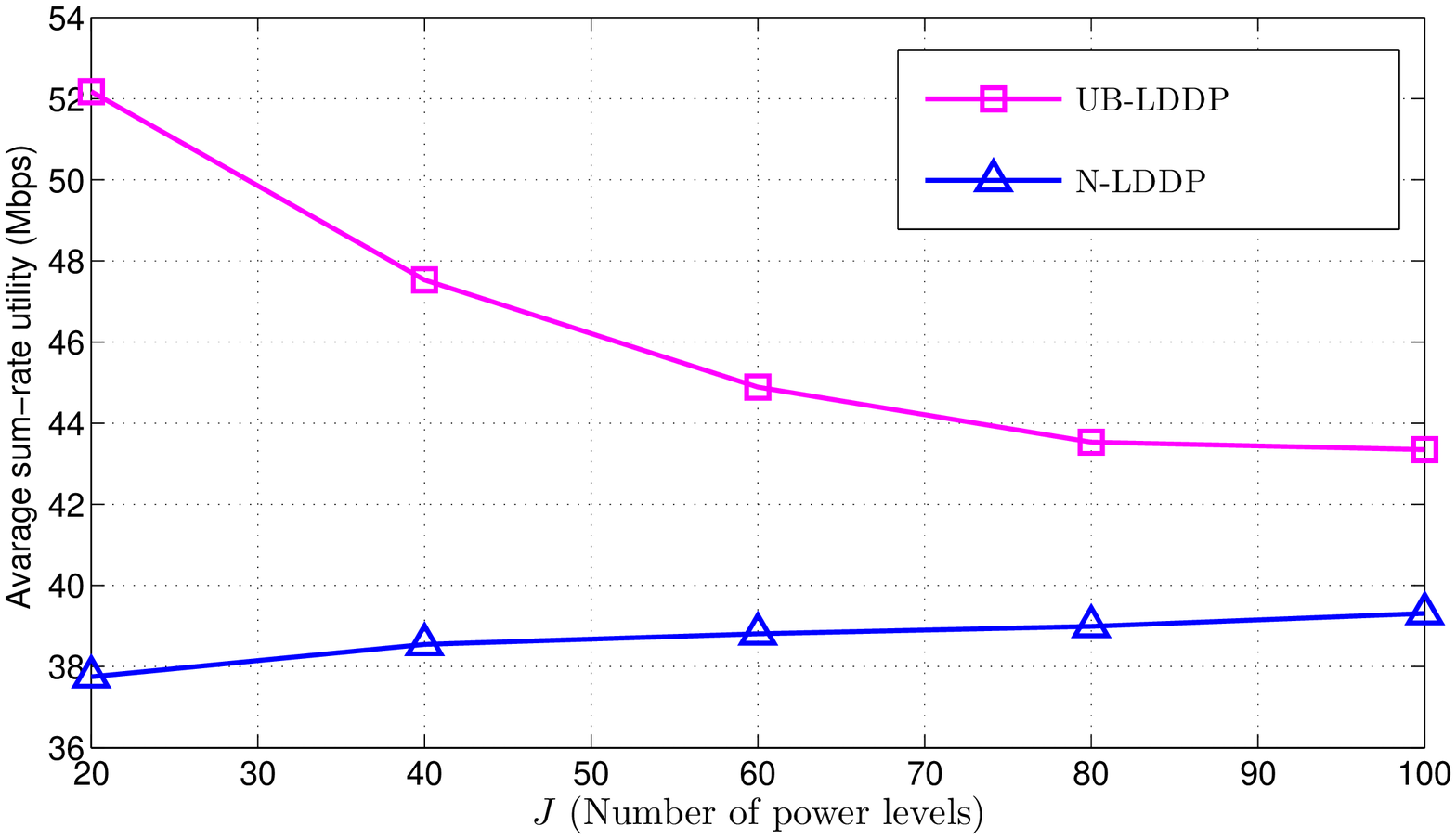}
\vspace{-1.3cm}
\caption{\ALG{} bounding performance in respect of parameter $J$.}
\label{fig:SRL}
\end{minipage}
\vspace{-0.8cm}
\end{figure}

\begin{figure}[htbp]
\begin{minipage}[t]{0.45\textwidth}
\centering
\includegraphics[scale=0.37]{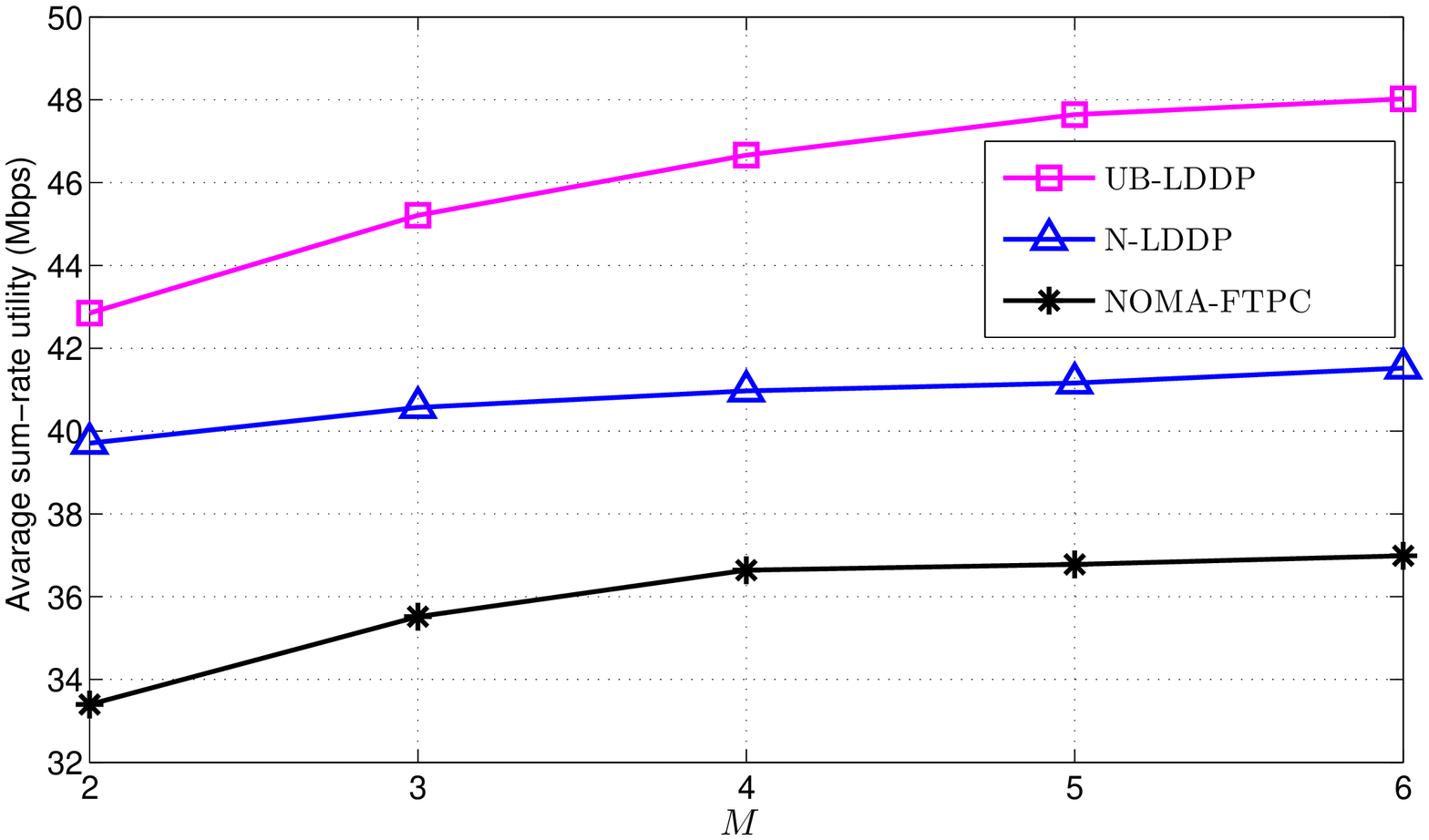}
\vspace{-1.3cm}
\caption{SR utility comparison in respect of parameter $M$.}
\label{fig:SRM}
\end{minipage}
\hspace{1.0cm}
\begin{minipage}[t]{0.45\textwidth}
\centering
\includegraphics[scale=0.37]{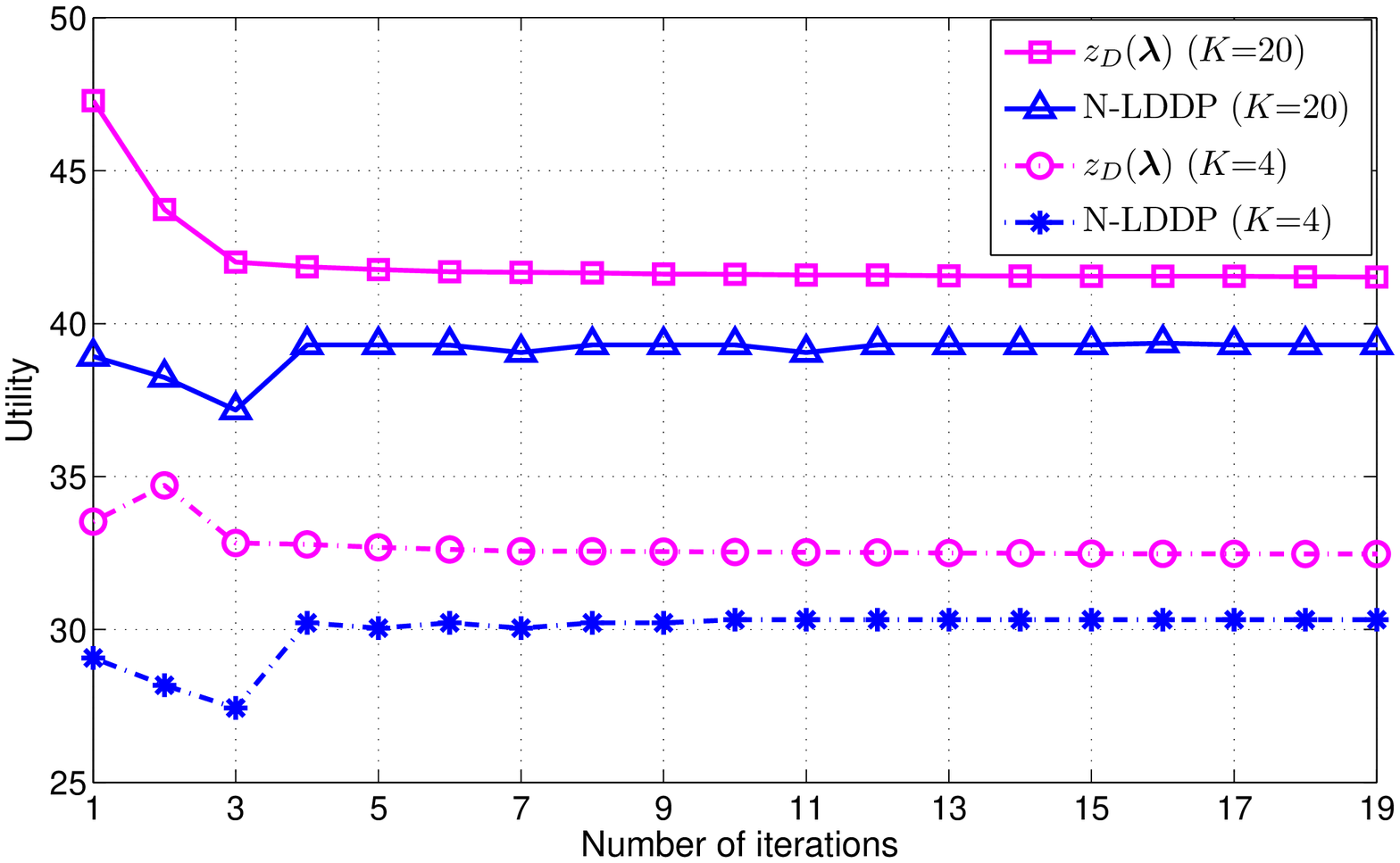}
\vspace{-1.3cm}
\caption{Illustration for \ALG{} convergence with $K=$ 4 and 20, $N=5$, $J=100$, and $M=2$.}
\label{fig:converge}
\end{minipage}
\vspace{-0.8cm}
\end{figure}

Applying the three algorithms to all instance, the average results are
summarized in Fig.~\ref{fig:SRuser} to \ref{fig:SRM}.  In
Fig.~\ref{fig:SRuser}, we evaluate the SR utility with respect
to user number $K$, with setting $M=2$ and $J=100$.  We make the
following observations.  First, the performance improvement tends to
be marginal for larger $K$ in all the schemes.  This is expected since
the multiuser diversity is effective when the number of users is
small, and is saturated if $K$ is large.  Second, N-\ALG{} outperforms
NOMA-FTPC and OFDMA-FTPC.  N-\ALG{} achieves performance improvement
of around 20\% over NOMA-FTPC.  NOMA-FTPC, in turn, performs much
better than OFDMA.  Third, N-\ALG{} is capable of providing
near-optimal solutions.  The average gap between UB-\ALG{} and
N-\ALG{} is $11\%$ in average, and the variation of the gap is
insensitive to the number of users.  This implies that the gap between
N-\ALG{} and the global optimum $z^{\dagger}$ is even smaller than
$11\%$.


\vspace{-0.1cm}

Next, the impact of parameters $J$ and $M$ is evaluated.  The
instances with $K=20$ are used in the simulations, and $M=2$ in
Fig. \ref{fig:SRL} and $J=100$ in Fig. \ref{fig:SRM}.  From
Fig. \ref{fig:SRL}, increasing $J$ leads to progressively tighter
intervals between UB-\ALG{} and N-\ALG{} since larger $J$ provides
better granularity in power discretization, and thus improves the
solution quality.  Moreover, we observe that the improvement comes
mainly from UB-\ALG{}.  This is because, for delivering the upper
bound, the objective value in $P_{\text{LR-D}}'$ has been
intentionally over-calculated by $\delta=\frac{P_{tot}}{J}$.  Compared
to using $\delta=0$, applying $\delta > 0$ in the over-calculation
results in an excess of utility in the objective.  This excess part is
clearly $J$-related, and is significantly reduced when $J$ is large.


In Fig.~\ref{fig:SRM}, more users are allowed to share the same
subcarrier.  Increasing $M$ leads to more total throughput for both
NOMA schemes.  One can observe that N-\ALG{} constantly outperforms
NOMA-FTPC.  We also notice that increasing $M$ results in degradation
of UB-\ALG{}.  The main reason is that when $M$ grows, the
over-calculations in the objective of $P_{\text{LR-D}}'$ are
accumulated over all multiplexed users.


\vspace{-0.5cm}

\subsection{Performance in Convergence}

We illustrate the convergence behavior of \ALG{} by two representative
instances with 4 and 20 users, respectively, with $N=5$, $J=100$, and
$M=2$.  The evolution of the values of $z_D(\bm{\lambda})$ and
$V_{\text{LB}}$ over the iteration number $C$ is provided in
Fig.~\ref{fig:converge}.

From the figure, a majority of the iterations is part of the
tailing-off effect. The utility ($V_\text{LB}$ of the algorithm) and the Lagrangian dual function ($z_D(\bm{\lambda}$)) both
approach the achievable values with 10 iterations or fewer. Note that
each iteration has polynomial-time complexity, see Theorem~\ref{dp}.


\subsection{Performance in Fairness}

As our next part of results, we evaluate the performance for \ALG{},
NOMA-FTPC, and OFDMA-FTPC in fairness.  We examine the fairness over a
scheduling time period. As was mentioned earlier, scheduling in the
time domain is slotted. Denote by $t$ the time slot index. We define a
scheduling frame consisting of $20$ slots. The channel state
information is collected once per frame.  Define $\bar
R_k(t)=(1-\frac{1}{T}) \bar R_k(t-1)+\frac{1}{T}r_k(t-1)$ as user $k$'s
average rate prior to slot $t$, where parameter $T$ is the length of a
time window (in the number of time slots), and $r_k(t-1)$ is user
$k$'s instantaneous rate in slot $t-1$ \cite{uplinksic, iccs}.  By
proportional fairness, the weight of user $k$ in slot $t$ is set to
$1/\bar R_k(t)$, and for each slot, N-\ALG{}, NOMA-FTPC, and
OFDMA-FTPC with WSR utility maximization are performed once.

Suppose the average users' rates are $\bar R_1,\dots, \bar R_K$ at the
end of the scheduling period, and consider the Jain's fairness index,
computed as $\frac{(\sum_{k=1}^{K}\bar R_k)^2}{K\sum_{k=1}^K \bar
R_k^2}$. This index, developed in
\cite{Jain}, is widely used as a fairness measure for user throughput
in communications networks.  The value of this fairness index is
between $\frac{1}{K}$ and $1.0$.  A higher value indicates fairer
throughput distribution, and the maximum value of $1.0$ is reached if
and only if all users achieve exactly the same throughput. Note that
the use of this index, by itself, does not prevent a user from being
served with low throughput (or even zero throughput), which, however,
will most likely bring down the value of the index.  In our case, zero
or very low throughput of any user is avoided by the fact that the
weights in \WOP{} are set in accordance with proportionally fair
scheduling.  Hence, over time, the weight of a user will increase to
infinity, if the user keeps being allocated with zero throughput. For
further insights of user throughput, particularly throughput of
cell-edge users, please see Section~\ref{sec:edge}.

\begin{figure}[htbp]
\begin{minipage}[t]{0.45\textwidth}
\centering
\includegraphics[scale=0.38]{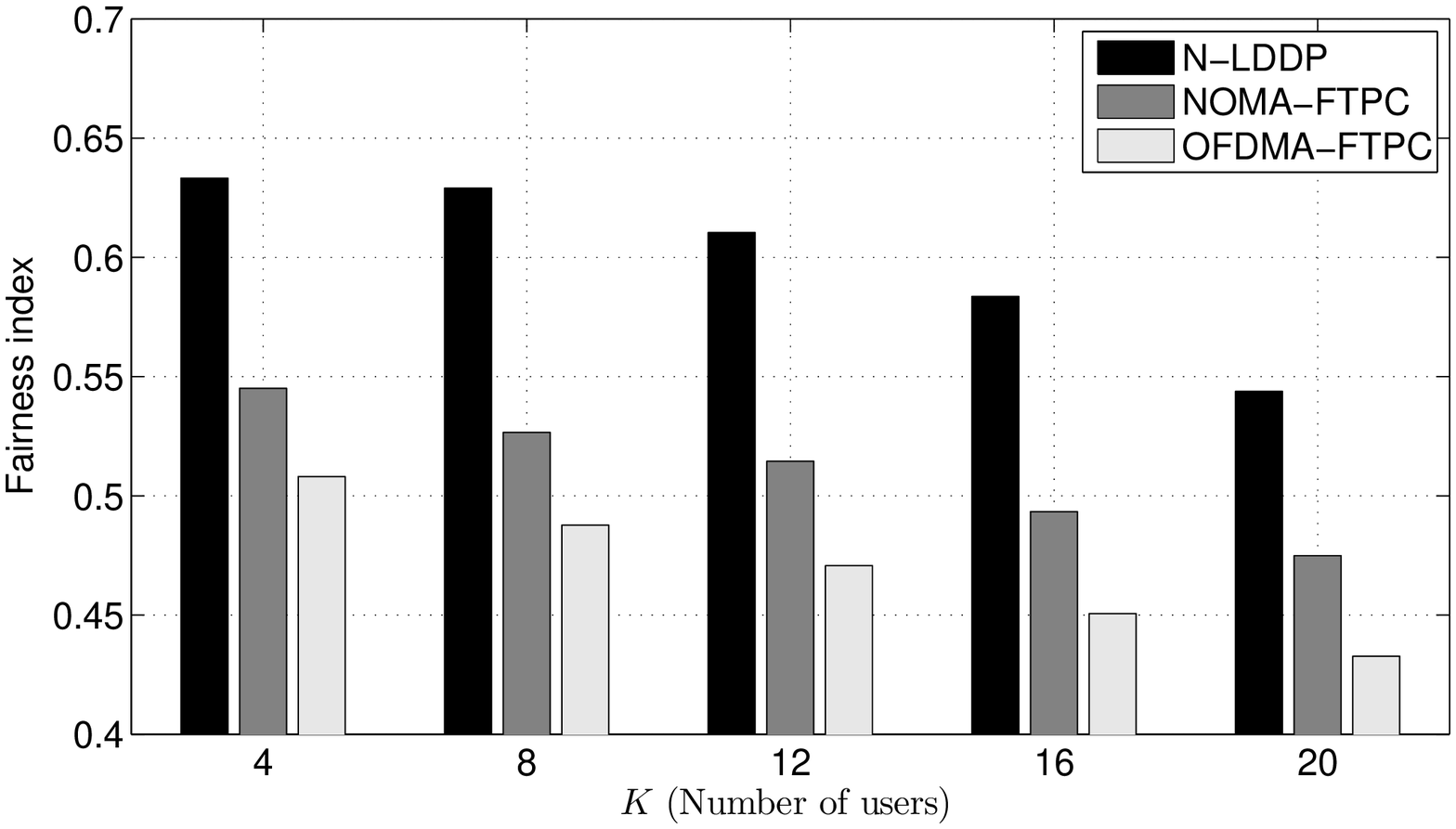}
\vspace{-1.3cm}
\caption{Fairness comparison in respect of the number of users.}
\label{fig:PFuser}
\end{minipage}
\hspace{1.0cm}
\begin{minipage}[t]{0.45\textwidth}
\centering
\includegraphics[scale=0.38]{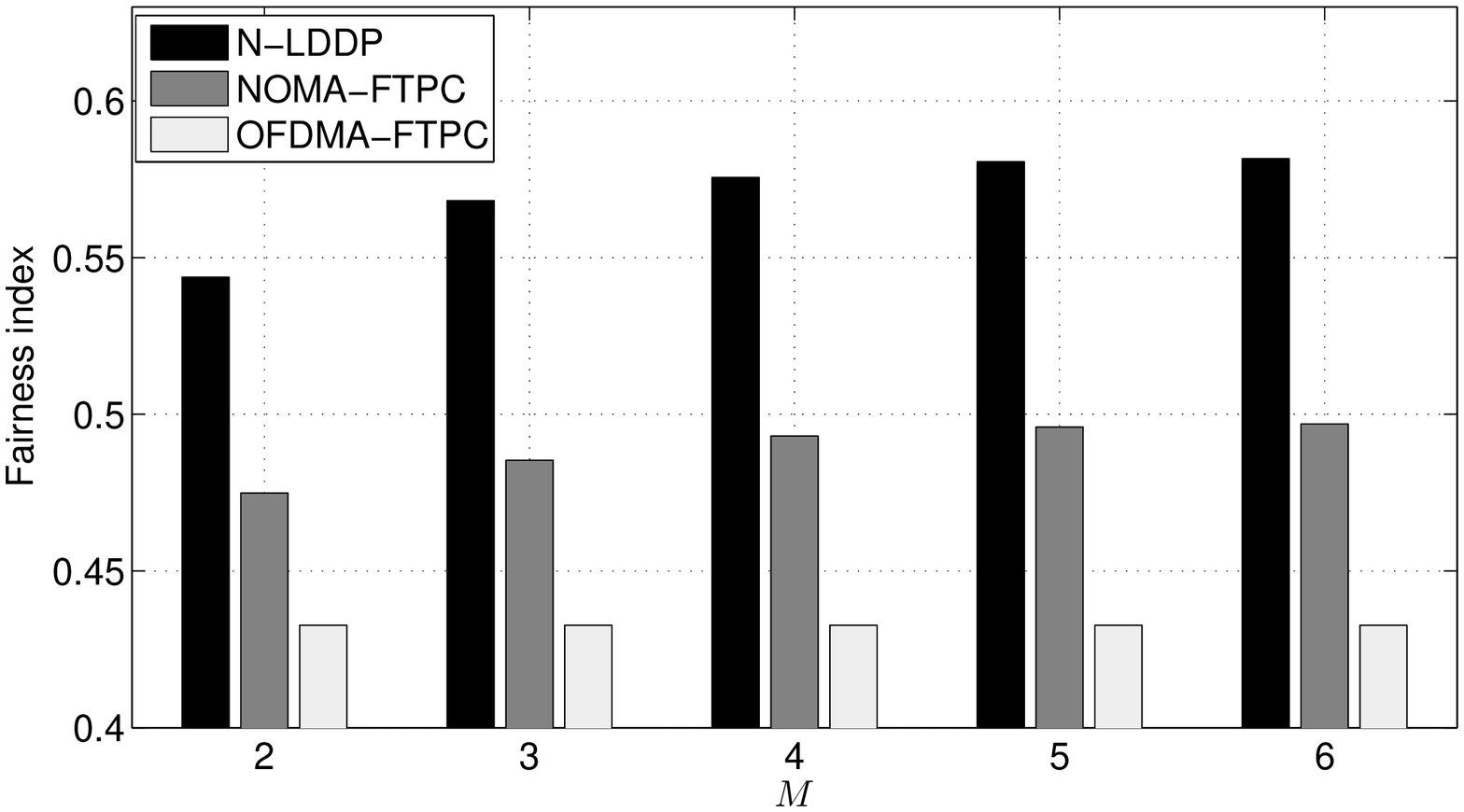}
\vspace{-1.3cm}
\caption{Fairness comparison in respect of parameter $M$.}
\label{fig:PFM}
\end{minipage}
\vspace{-0.5cm}
\end{figure}

The fairness index in respect of $K$ and $M$ is shown in
Fig. \ref{fig:PFuser} and Fig.~\ref{fig:PFM}, respectively.  The
parameters are set to be $M=2$ in Fig.~\ref{fig:PFuser} and $K=20$ in
Fig. \ref{fig:PFM}.  For both figures, we consider a scheduling period
of $100$ time slots, with $J=100$ and $T=50$.  From
Fig.~\ref{fig:PFuser} and \ref{fig:PFM}, we observe that, first, the
proposed N-\ALG{} achieves the best performance.  Moreover,
Fig.~\ref{fig:PFuser}, increasing $K$ leads to fairness degradation in
all schemes due to more competition among the users.

%

\vspace{-0.5cm}

From Fig.~\ref{fig:PFM}, the fairness index increases in $M$. This is
because a larger $M$ provides more flexibility in resource allocation
among the users.  There is however a saturation effect, showing that
the constraining impact due to limiting the number of
multiplexed users per subcarrier decreases when the limit becomes
large.  Note that in Fig.~\ref{fig:SRM}, the improvement of throughput
is marginal with $M$ becomes large. Thus a moderate $M$ is justified
not only by implementation complexity, but also that having a large
$M$ may not lead to significant performance improvement.

In the two figures, OFDMA-FTPC gives the lowest fairness index. A particular
reason is that the FTPC channel and  power
allocation scheme is sub-optimal. This also explains the improvement
enabled by the proposed power optimization algorithm in comparison to
NOMA-FTPC. We also remark that the vertical axis of the two figures
starts from a positive value, hence the relative difference between
the schemes in fairness is smaller than what it may appear to be.

\subsection{Performance for Cell-edge Users in Throughput}
\label{sec:edge}

\begin{figure}[htbp]
\begin{minipage}[t]{0.45\textwidth}
\centering
\includegraphics[scale=0.38]{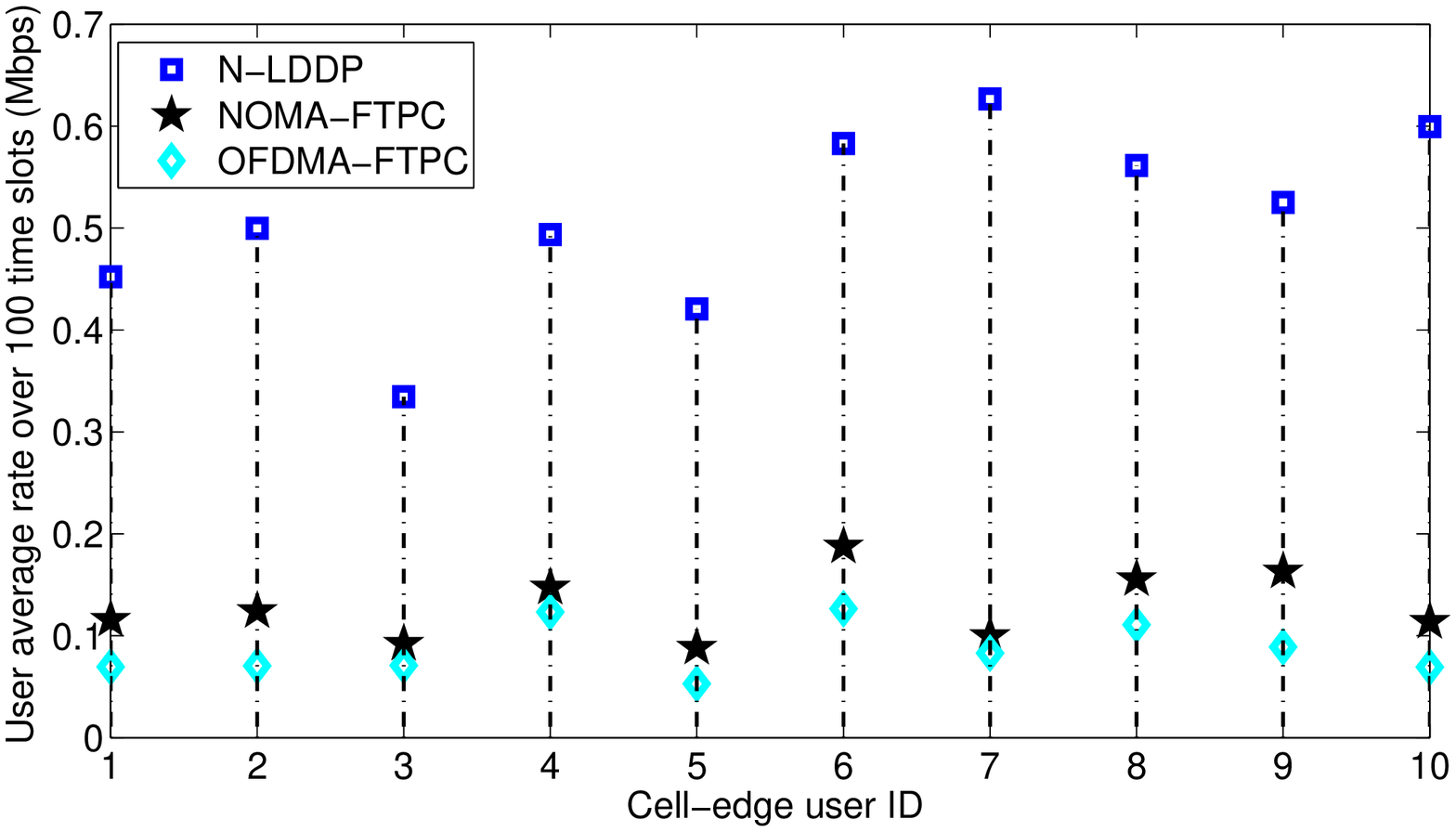}
\vspace{-1.3cm}
\caption{Performance comparison in cell-edge users.}
\label{fig:PFedge}
\end{minipage}
\hspace{1.0cm}
\begin{minipage}[t]{0.45\textwidth}
\centering
\includegraphics[scale=0.39]{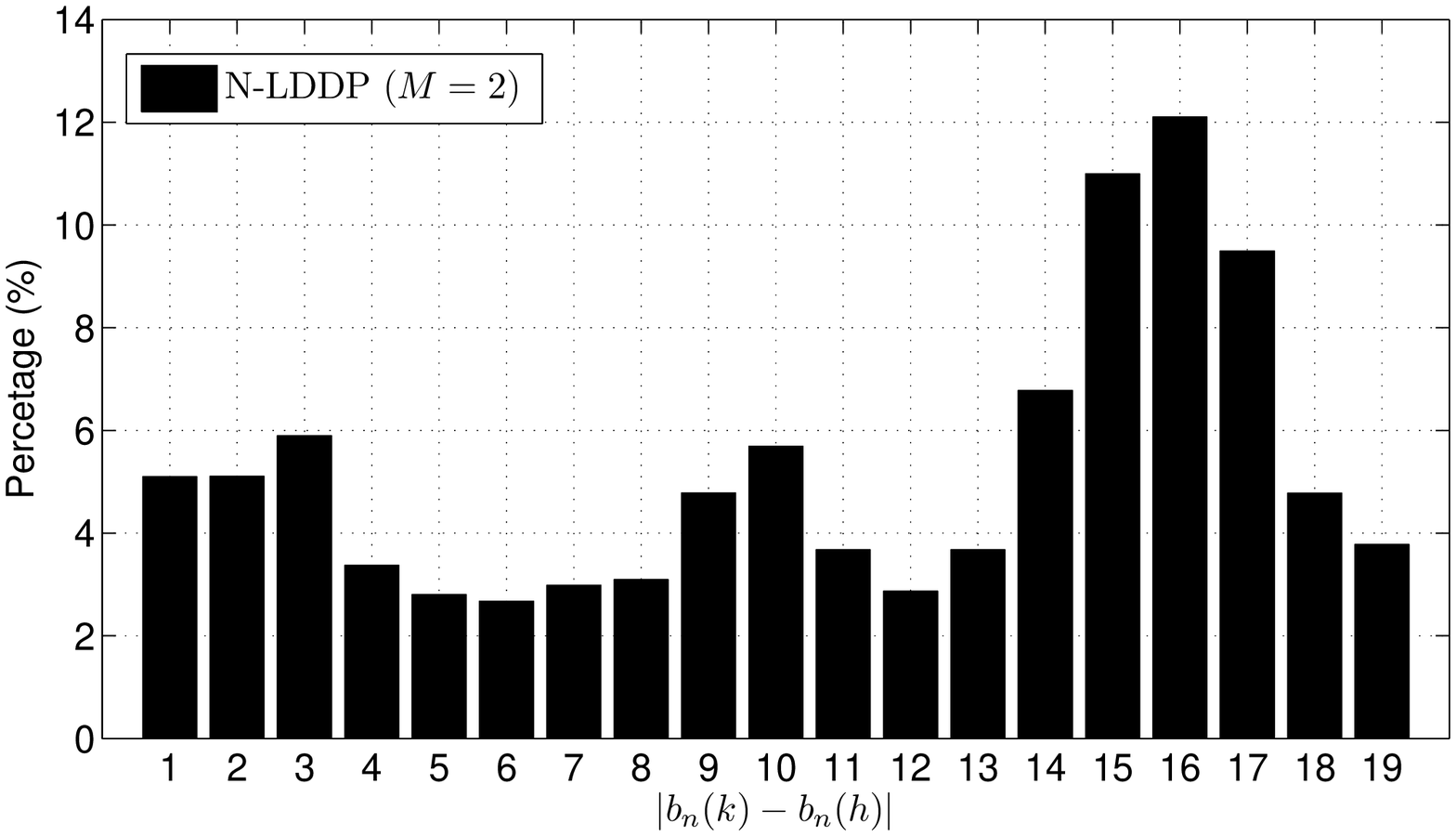}
\vspace{-1.3cm}
\caption{The results of user grouping in N-\ALG{}.}
\label{fig:order}
\end{minipage}
\vspace{-0.7cm}
\end{figure}

The performance of cell-edge users is of significance.  To evaluate,
we split the service area of the cell into an edge zone and a center
zone.  Performance comparison is carried out for twenty-user instances
with $M=2$, $J=100$, and 100 time slots.  In
the simulations, we deploy half of the users to the cell edge in each
instance.  Each value in Fig. \ref{fig:PFedge} represents the average
user rate over the entire scheduling period.  We observe that
using \ALG{} significantly improves the rates for cell-edge users.
From the results in Fig. \ref{fig:PFedge}, the average rates of all
cell-edge users in N-\ALG{} are much more than those of
NOMA-FTPC and OFDMA-FTPC.


\vspace{-0.5cm}

\subsection{Characteristics of User Grouping}

The final part of our performance study is on
characteristics of user grouping, i.e., which users tend to be
multiplexed together on the same subcarrier by optimization, We
consider a scenario with $M=2$, $J=100$, $K=20$, and apply N-\ALG{} to
1,000 realizations.  For each subcarrier, we index the users in
descending order of channel gain.  If there are two users
multiplexed on the subcarrier in the algorithm solution, we consider
the difference of the two user indices. For example, if the users with
the highest and lowest gains are grouped together, the difference is
19. The results are illustrated in Fig. \ref{fig:order}, where the
horizontal axis is the difference in index.


From the figure, we observe that users having large difference in
channel gain are more likely to be multiplexed on the same subcarrier.
This is coherent with the conclusion in \cite{DiFaPo15}. On the other
hand, it is also evident by the figure that optimal assignment is not
necessarily to select users with the best and poorest gains on a
subcarrier.  The observation motivates the treatment of subcarrier
allocation as optimization variables.

\section{Conclusions}

We have considered jointly optimizing power and channel allocation for
NOMA.  Theoretical insights on complexity and optimality have been
provided, and we have proposed an algorithm framework based on
Lagrangian dual optimization and dynamic programming.  The proposed
algorithm is capable of providing near-optimal solutions as well as
bounding the global optimum tightly.  Numerical results demonstrate
that the proposed algorithmic notions result in significant
improvement of throughput and fairness in comparison to existing OFDMA
and NOMA schemes.

An extension of the work is the consideration of max-min fairness for
one scheduling instance. In this case, one solution approach is to
perform a bi-section search.  For each target level that represents
the minimum throughput required for all users, the problem reduces to
a feasibility test, i.e., whether or not the target is achievable
subject to the power limits. This can be formulated as to minimize the
total power, with constraints specifying the throughput target value
and user-individual power limits. The development of optimization
algorithms for this problem is subject to further study.

\section*{ACKNOWLEDGEMENTS}

We would like to thank the anonymous reviewers for their comments and
suggestions. This work has been supported by the European Union Marie
Curie project MESH-WISE (324515), Career LTE (329313). The work of the
first author has been supported by the China Scholarship Council
(CSC).


\end{document}